\documentclass[useAMS,usenatbib]{mnras}

\usepackage{graphicx, bm, amssymb, xcolor}
\usepackage{verbatim}
\usepackage[all]{hypcap}
\usepackage{xspace}
\usepackage{multirow,bigdelim}
\usepackage[fleqn]{amsmath}

\usepackage{epsfig}
\usepackage{aas_macros}
\usepackage{natbib}
\usepackage{times,txfonts}
\usepackage{morefloats}
\usepackage{tabularx}
\usepackage{lscape}
\usepackage{makecell}

\newcommand{\msun}{\mbox{M$_\odot$}}
\newcommand{\yr}{\mbox{${\rm yr}$}}

\newcommand{\gyr}{\mbox{${\rm Gyr}$}}

\newcommand{\kpc}{\mbox{${\rm kpc}$}}

\newcommand{\dex}{\mbox{${\rm dex}$}}
\newcommand{\feh}{\mbox{$[{\rm Fe}/{\rm H}]$}}

\newcommand{\mh}{\mbox{$M_{\rm h}$}}
\newcommand{\ms}{\mbox{$M_\star$}}
\newcommand{\ra}{\mbox{$R_{\rm a}$}}
\newcommand{\ecc}{\mbox{$\epsilon$}}
\newcommand{\ngc}{\mbox{$N_{\rm GC}$}}
\newcommand{\mgcs}{\mbox{$M_{\rm GCS}$}}
\newcommand{\zacc}{\mbox{$z_{\rm acc}$}}
\newcommand{\tacc}{\mbox{$t_{\rm acc}$}}
\newcommand{\tn}{\mbox{$T_{\rm N}$}}
\newcommand{\etan}{\mbox{$\eta_{\rm N}$}}
\newcommand{\etam}{\mbox{$\eta_{\rm M}$}}
\newcommand{\rmass}{\mbox{$r_{M_\star}$}}
\newcommand{\rhmass}{\mbox{$r_{M_{\rm h}}$}}
\newcommand{\nbrz}{\mbox{$N_{{\rm br},z>2}$}}
\newcommand{\nbr}{\mbox{$N_{\rm br}$}}

\newcommand{\lcdm}{$\Lambda$CDM\xspace}
\newcommand{\mosaics}{MOSAICS\xspace}
\newcommand{\emosaics}{E-MOSAICS\xspace}
\newcommand{\eagle}{EAGLE\xspace}
\newcommand{\ges}{{\it Gaia}-Enceladus\xspace}

\newcommand{\python}{{\sc Python}\xspace}
\newcommand{\skl}{{\sc scikit-learn}\xspace}
\newcommand{\keras}{{\sc Keras}\xspace}
\newcommand{\be}{\begin{equation}}
\newcommand{\ee}{\end{equation}}
\newcommand{\bea}{\begin{eqnarray}}
\newcommand{\eea}{\end{eqnarray}}

\usepackage{etoolbox}
\makeatletter
\patchcmd\@combinedblfloats{\box\@outputbox}{\unvbox\@outputbox}{}{%
    \errmessage{\noexpand\@combinedblfloats could not be patched}%
}%
\makeatother

\newcommand{\appropto}{\mathrel{\vcenter{
  \offinterlineskip\halign{\hfil$##$\cr
    \propto\cr\noalign{\kern1pt}\sim\cr\noalign{\kern-2pt}}}}}

\title[Kraken reveals itself]{\vspace{-5mm}Kraken reveals itself -- the merger history of the Milky Way reconstructed with the E-MOSAICS simulations\vspace{-4mm}}

\author{J.~M.~Diederik Kruijssen,$^{1,2}$\thanks{E-mail: \href{kruijssen@uni-heidelberg.de}{kruijssen@uni-heidelberg.de}}
Joel~L.~Pfeffer,$^3$
M\'{e}lanie~Chevance,$^1$
Ana Bonaca,$^2$
\newauthor
Sebastian Trujillo-Gomez,$^1$
Nate~Bastian,$^3$ 
Marta~Reina-Campos,$^1$
Robert~A.~Crain,$^3$
\newauthor
and Meghan~E.~Hughes$^3$ \\
$^1$Astronomisches Rechen-Institut, Zentrum f\"{u}r Astronomie der Universit\"{a}t Heidelberg, M\"{o}nchhofstra\ss e 12-14, 69120 Heidelberg, Germany\\
$^2$Institute for Theory and Computation, Harvard University, 60 Garden Street, Cambridge, MA 02138, USA\\
$^3$Astrophysics Research Institute, Liverpool John Moores University, IC2, Liverpool Science Park, 146 Brownlow Hill,
Liverpool L3 5RF, UK\vspace{-3mm}}

\begin{document}

\date{Accepted X{\sevensize xxxx} XX. Received X{\sevensize xxxx} XX; in original form 2020 March 1\vspace{-3mm}}

\pagerange{\pageref{firstpage}--\pageref{lastpage}} \pubyear{2020}

\maketitle

\label{firstpage}

\begin{abstract}
Globular clusters (GCs) formed when the Milky Way experienced a phase of rapid assembly. We use the wealth of information contained in the Galactic GC population to quantify the properties of the satellite galaxies from which the Milky Way assembled. To achieve this, we train an artificial neural network on the \emosaics cosmological simulations of the co-formation and co-evolution of GCs and their host galaxies. The network uses the ages, metallicities, and orbital properties of GCs that formed in the same progenitor galaxies to predict the stellar masses and accretion redshifts of these progenitors. We apply the network to Galactic GCs associated with five progenitors: \ges, the Helmi streams, Sequoia, Sagittarius, and the recently discovered, `low-energy' GCs, which provide an excellent match to the predicted properties of the enigmatic galaxy `Kraken'. The five galaxies cover a narrow stellar mass range [$\ms=(0.6{-}4.6)\times10^8~\msun$], but have widely different accretion redshifts ($\zacc=0.57{-}2.65$). All accretion events represent minor mergers, but Kraken likely represents the most major merger ever experienced by the Milky Way, with stellar and virial mass ratios of $\rmass=1$:$31^{+34}_{-16}$ and $\rhmass=1$:$7^{+4}_{-2}$, respectively. The progenitors match the $z=0$ relation between GC number and halo virial mass, but have elevated specific frequencies, suggesting an evolution with redshift. Even though these progenitors likely were the Milky Way's most massive accretion events, they contributed a total mass of only $\log{(M_{\rm \star,tot}/\msun)}=9.0\pm0.1$, similar to the stellar halo. This implies that the Milky Way grew its stellar mass mostly by in-situ star formation. We conclude by organising these accretion events into the most detailed reconstruction to date of the Milky Way's merger tree.
\end{abstract}

\begin{keywords}
galaxies: evolution --- galaxies: formation --- galaxies: haloes --- galaxies: star formation --- globular clusters: general --- Galaxy: formation\vspace{-4mm}
\end{keywords}

\section{Introduction} \label{sec:intro}

It is one of the main goals in modern galaxy formation studies to reconstruct and understand the assembly histories of galaxies \citep[e.g.][]{eggen62,searle78,ibata94,belokurov06,bell08,johnston08,mcconnachie09,cooper10,deason13,pillepich14,kruijssen19d}. Specifically, satellite galaxy accretion histories, often expressed in terms of merger trees, represent a clear and testable prediction of structure formation in the cold dark matter (\lcdm) cosmology \citep[e.g.][]{bullock05,deason15,fattahi20}. In order to reconstruct these accretion histories, it is necessary to obtain a comprehensive census of the redshifts at which satellites were accreted and the (stellar or halo\footnote{We use the terms `halo mass' and `virial mass' to refer to the sum of the dark matter and baryonic mass of the galaxy within its virial radius.}) masses of these systems at the time of accretion. This can be done by identifying a set of observables that traces the galaxy merger tree of the host galaxy. In the Milky Way, this has recently become possible thanks to two major developments. First, the {\it Gaia} satellite has provided near-complete six-dimensional (position-velocity) phase space information for an unprecedented number of stars and stellar clusters in the Milky Way \citep[e.g.][]{gaia18b,baumgardt19,vasiliev19}, which together provides the potential means of inferring the accretion histories of a wide variety of satellite progenitors. Secondly, the modelling frameworks have recently been developed to connect the observed phase space (e.g.\ orbital) information to the properties of the progenitor satellites \citep[e.g.][]{belokurov18,haywood18,helmi18,myeong18,koppelman19,kruijssen19d,kruijssen19e,massari19,helmi20}.

In particular, the use of globular clusters (GCs) to trace the assembly history of the Milky Way has seen an increase in applications \citep[e.g.][]{forbes18,helmi18,myeong18,myeong19,kruijssen19e}. Various combinations of GC energies and angular momenta (i.e.\ orbits and integrals of motion, \citealt{myeong18,helmi18}), as well as GC ages and metallicities \citep[e.g.][]{forbes10,leaman13,li14,choksi18,kruijssen19e} have provided important constraints on the satellite population that was accreted by the Milky Way. Most recently, these efforts have been aided by the \emosaics project, which is a suite of self-consistent, hydrodynamical cosmological simulations with a complete model for the formation and evolution of the GC population \citep{pfeffer18,kruijssen19d}. Because these simulations simultaneously reproduce young and old stellar cluster populations with a single, environmentally dependent cluster formation and disruption model, they enable linking the properties of the cluster population to the assembly history of the host galaxy.

In a recent paper, we used the age-metallicity distribution of Galactic GCs to infer the formation and assembly history of the Milky Way, culminating in the partial reconstruction of its merger tree \citep{kruijssen19e}. This work made use of the correlation in the \emosaics simulations between quantities describing the formation assembly histories of galaxies (e.g.\ the dark matter halo concentration, the number of accretion events, the total number of progenitors, and the number of minor mergers) and the properties of their host GC populations (e.g.\ the number of GCs, the slope of the age-metallicity distribution, and the median age) to characterise the assembly history of the Milky Way, and additionally used the detailed distribution of Galactic GCs in age-metallicity space to derive the number of accreted satellites and their stellar mass growth histories. The main conclusions of that work are as follows.
\begin{enumerate}
\item
The Galactic GC age-metallicity distribution bifurcates into a steep, `main' branch of GCs that formed in-situ in the Main progenitor of the Milky Way and a shallow, `satellite' branch at lower metallicities that is constituted by accreted GCs that formed in low-mass satellite galaxies. A total of $\sim15$ such satellites must have been accreted based on the number of Galactic GCs and the slope of their age-metallicity distribution, even if only a minority of these satellites is expected to have brought in detectable numbers of GCs. At least some of these accretion events may have had no associated GCs at all \citep[also see e.g.][]{koppelman19b}.
\item
The steepness of the main branch implies that the Milky Way assembled quickly for its mass, reaching \{25, 50\}~per~cent of its present-day halo mass already at $z=\{3, 1.5\}$ and half of its present-day stellar mass at $z=1.2$. The growth history of the Milky Way runs ahead of those typical for galaxies of its $z=0$ mass \citep[e.g.][]{papovich15} by about 1~Gyr.
\item
There are too many GCs on the satellite branch to be attributable to a single progenitor, because the number of GCs found in this branch is considerably larger than expected for the low masses of the galaxies \citep[e.g.][]{harris13,harris17} forming stars at the low metallicities characterising the satellite branch. At least two (and preferably three) massive satellites are needed to contribute most of the GCs on the satellite branch, which is supported by the existence of multiple kinematic components that were known pre-{\it Gaia} (including the Sagittarius dwarf galaxy, \citealt{ibata94}). It is possible that the satellite branch contains traces of a larger number of satellites (e.g.\ those with only 1--2 GCs, which are not detectable due to Poisson noise), but the age-metallicity distribution does not provide sufficient constraints to tell apart small sub-groups. Kinematic information would be necessary to potentially lift this degeneracy.
\item
The satellite branch in age-metallicity space is quite narrow, with a total metallicity spread of $\Delta\feh\approx0.3~\dex$, implying that the masses of the $>3$ progenitor satellites were similar at any given lookback time or redshift. For this reason, \citet{kruijssen19e} do not distinguish between the two most massive satellites, implying that their masses were similar to within a factor of $\sim2$.
\item
Out of the three identified satellite progenitors, the most recent (and at a given lookback time least massive) accretion event corresponds to Sagittarius \citep{ibata94}. The next most massive satellite brought in a large number of GCs, many of which were formerly associated with the Canis Major `mirage', a perceived accretion event that never existed and rather represents a density wave in the Galactic disc \citep[e.g.][]{martin04,penarrubia05,deason18,deboer17}. The actual satellite that brought in these GCs has since been found in the {\it Gaia} data \citep{belokurov18} and was dubbed the {\it Gaia} Sausage \citep{myeong18} or \ges \citep{helmi18}.
\item
The third progenitor was dubbed `Kraken' and must have had a mass very similar to that of {\it Gaia}-Enceladus. Until recently, it had not been found. However, in a recent paper, \citet{massari19} identified a group of GCs at low energies in the {\it Gaia} DR2 data. These GCs represent a significant fraction of the satellite branch in age-metallicity space and thus likely represent the Kraken progenitor event needed to explain the age-metallicity `satellite branch' GCs after Sagittarius and {\it Gaia}-Enceladus have been accounted for.\footnote{\citet{massari19} caution that the GCs suggested by \citet{kruijssen19e} to have been potential members of Kraken do not quite match their kinematic selection. However, \citet{kruijssen19e} used kinematic information from the literature that preceded {\it Gaia} DR2 and therefore explicitly refrained from associating individual GCs with any particular satellites. Instead, we encouraged future studies to look for phase-space correlations between the suggested sets of GCs and those that have been proposed to be associated with particular accretion events. As such, the proposed potential member GCs merely represented `wish lists' of interesting targets for kinematic follow-up work rather than definitive member lists. As a result, the existence (or not) of Kraken cannot be evaluated based on the possible membership of individual GCs proposed by \citet{kruijssen19e}.} It is one of the main goals of this paper to determine the mass and accretion redshift of the progenitor that brought in the `low-energy' GCs from \citet{massari19} and thus assess whether this progenitor is Kraken.
\end{enumerate}

In addition to Kraken, \ges, and Sagittarius, \citet{massari19} find that the satellite branch accommodates GCs from two other accreted satellites, i.e.\ the progenitor of the `Helmi streams' \citep{helmi99} and `Sequoia' \citep{myeong19}. In a recent paper, \citet{forbes20} used the numbers of GCs that \citet{massari19} assign to each of the five satellite progenitors to estimate the galaxy masses, confirming our interpretation that the low-energy GCs match the predicted properties of Kraken. However, as we discuss in Section~\ref{sec:ms}, the approach of using present-day GC numbers to estimate the host galaxy mass at accretion systematically overestimates the galaxy masses by up to a factor of 3.

In this paper, we use the groups of GCs identified by \citet{massari19} as having been accreted from the same satellite progenitors to determine the masses and accretion redshifts of these five galaxies. To do so, we train an artificial neural network to connect the properties of the GCs contributed by individual accreted satellites in the \emosaics simulations to the properties of their host accretion events. Specifically, we use the median and interquartile ranges of the GC apocentre radii, eccentricities, ages, and metallicities as feature variables to predict the target variables of the accretion redshift and the host stellar mass at the time of accretion. The resulting neural network is then applied to the groups of GCs identified by \citet{massari19}. By combining the resulting predictions with the constraints on the assembly history of the Milky Way from \citet{kruijssen19e}, we infer the merger tree of the Milky Way, identifying five specific accretion events. Future work applying a similar methodology to field stars is expected to extend this analysis to additional satellite progenitors.

The structure of this paper is as follows. In Section~\ref{sec:nn}, we describe the procedure followed to train the neural network on the \emosaics simulations. In Section~\ref{sec:prog}, we present the accretion redshifts and stellar masses of the satellite progenitors, and compare the resulting properties of the progenitors to scaling relations describing galaxies and their GC populations in the nearby Universe. In Section~\ref{sec:hist}, we combine this with the stellar mass growth history of the Milky Way to determine the stellar mass ratios of the accretion events and reconstruct the merger tree of the Milky Way. We present our conclusions in Section~\ref{sec:concl}.

\section{An artificial neural network predicting satellite masses and accretion redshifts from GC demographics} \label{sec:nn}

\subsection{Simulation suite and training set} \label{sec:sims}
In this paper, we use the suite of 25 zoom-in simulations from the \emosaics project \citep{pfeffer18,kruijssen19d} to provide a training set based on which the properties of the progenitor galaxies of groups of GCs can be predicted. \emosaics is a suite of hydrodynamical cosmological simulations with a complete, self-consistent model for the formation and evolution of the GC population. This suite of simulations combines the model for galaxy formation and evolution from \eagle \citep{crain15,schaye15} with the sub-grid model for the formation and evolution of the entire stellar cluster population \mosaics \citep{kruijssen11,kruijssen12c,pfeffer18}. All simulations adopt a $\Lambda$CDM cosmogony, described by the parameters advocated by the \citet{planck14}, namely $\Omega_0 = 0.307$, $\Omega_{\rm b} = 0.04825$, $\Omega_\Lambda= 0.693$, $\sigma_8 = 0.8288$, $n_{\rm s} = 0.9611$, $h = 0.6777$, and $Y = 0.248$. 

\emosaics reproduces the demographics of young cluster populations in nearby galaxies \citep{pfeffer19b} \citep[as well as predicts those of high-redshift galaxies, see][]{pfeffer19,reinacampos19} and simultaneously reproduces a wide variety of observables describing the old GC population in the local Universe, such as the number of GCs per unit galaxy mass, radial GC population profiles, and the high-mass ($M>10^5~\msun$) end of the GC mass function \citep{kruijssen19d}, as well as the mass-metallicity relation of metal-poor GCs \citep[the `blue tilt'][also see \citealt{kruijssen19c}]{usher18}, the GC age-metallicity distribution \citep{kruijssen19e}, the kinematics of GC populations \citep{trujillogomez20}, the dynamical mass loss histories of massive GCs \citep{reinacampos18,reinacampos20,hughes20}, and the association of GCs with fossil stellar streams from accreted dwarf galaxies \citep{hughes19}.

The goal of this work is to predict the accretion redshifts ($\zacc$) and the stellar masses ($\ms$) at the time of accretion onto the Milky Way of the satellite progenitors identified by \citet{massari19}. To this end, we train an artificial neural network with `target' variables $\zacc$ and $\ms$ as a function of eight `feature' variables that describe the properties of the GCs associated with each of these progenitors. In \emosaics, the accretion redshift is defined as the moment at which {\sc subfind} \citep{springel01,dolag09} can no longer find a bound subhalo and the subhalo is therefore considered to have merged into the halo of the central halo (see \citealt{qu17} for discussion). As feature variables, we use the medians and interquartile ranges (IQRs) of the GC age ($\tau$), GC metallicity ($\feh$), GC orbital apocentre radius ($\ra$), and GC orbital eccentricity ($\ecc$). Across all 25 simulations, we identify all progenitor satellites with stellar masses $\log{(\ms/\msun)}\geq6.5$ that host GCs and are accreted onto the central galaxy, for a total of $N_{\rm sat}=205$ accretion events, or $\sim8$ per Milky Way-mass galaxy on average. For each of these satellite progenitors, we tabulate the target variables describing the accretion event and the feature variables describing the properties of the sub-populations of GCs contributed by these satellites. Because \emosaics overpredicts the number of GCs at masses much smaller than $M\sim10^5~\msun$ due to underdisruption \citep{pfeffer18,kruijssen19d}, we only consider GCs with $z=0$ masses of $M\geq5\times10^4~\msun$. We verified that the exact choice of this lower mass limit does not strongly affect the results of this work, because the median and IQR of GC ages, metallicities, apocentre radii, and eccentricities are not strongly correlated with the GC mass. The adopted limit of $M\geq5\times10^4~\msun$ is found to provide the best compromise between minimising the effects of underdisruption (requiring high GC masses) and having a sufficient number of GCs (requiring low GC masses). In addition, we restrict the GC metallicities to $-2.5<\feh<-0.5$, in order to match the range of metallicities of Galactic GCs for which age measurements are available \citep{marinfranch09,forbes10,dotter10,dotter11,leaman13,kruijssen19e}.

Of course, the \emosaics simulations themselves only provide a finite (and quite small) sample of 25 Milky Way-mass galaxies and their accretion histories. There is clear evidence that the accretion history of the Milky Way is atypical for a galaxy of its mass \citep[e.g.][]{deason16,belokurov18,kruijssen19e,mackereth19}. Ideally, we would therefore have been able to use many more simulations, in order to ensure that the intricacies of the Milky Way's particular accretion history are captured by at least one of the galaxies in the sample. However, we do not find strong variations in the relations between feature and target variables across the suite of simulations \citep[see Section~\ref{sec:nndetails} and][]{pfeffer20}, which gives some confidence that the application of these relations in this paper is robust.

\subsection{Details of the neural network} \label{sec:nndetails}
\begin{figure*}
\includegraphics[width=\hsize]{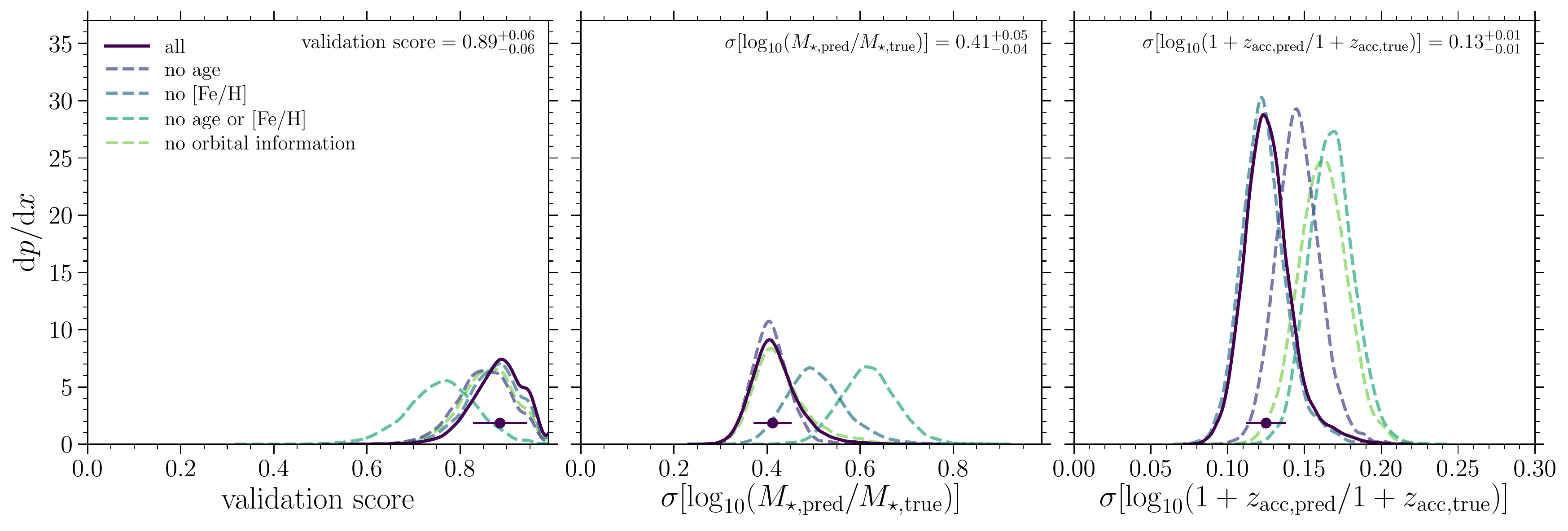}%
\caption{
\label{fig:f1}
Validation of the neural network. Left: PDF of validation scores across all 10,000 Monte Carlo realisations of the neural network, for models including different subsets of the feature variables as indicated by the legend. The data point with error bar and the annotation in the top right indicate the median and 16th-to-84th percentiles for the experiment including all feature variables. Middle: PDF of the logarithmic scatter of the predicted stellar masses of the satellite progenitors around their true stellar masses. Lines and annotation as in the left-hand panel. Right: PDF of the logarithmic scatter of the predicted accretion redshifts of the satellite progenitors around their true accretion redshifts. Lines and annotation as in the left-hand panel. This figure shows that the neural network performs well, predicting stellar masses ($\ms$) within a factor of 2.5 and accretion redshifts ($1+\zacc$) within a factor of 1.35.}
\end{figure*}	
We train a sequential neural network to predict the accretion redshifts and stellar masses of galaxies accreted by the Milky Way using the \python packages \skl \citep{pedregosa11,buitinck13} and \keras \citep{chollet15}, which are application programming interfaces for machine learning and neural networks, respectively. We construct a neural network with `dense' hidden layers, i.e.\ one in which all nodes in successive layers are connected. The hidden layers use a Rectified Linear Unit ({\tt ReLU}) activation function, i.e.\ $f(x)=\max{(0,x)}$. The architecture of the network is chosen by optimising the validation score while varying the number of hidden layers in the range $N_{\rm lay}=1{-}10$ and the number of nodes per layer in the range $N_{\rm node}=10{-}80$. The validation score shows little variation across the parameter space probed, with a slight maximum around $N_{\rm lay}=4$ and $N_{\rm node}=50$ (for which the validation score is $0.89^{+0.06}_{-0.06}$, see below). While we adopt these numbers throughout this work, changing them by up to a factor of 2 does not qualitatively affect our results. The hidden layers are connected to an input layer and an output layer. By definition, the input layer consists of 8 nodes (reflecting the number of feature variables) and the output layer consists of 2 nodes (reflecting the number of target variables). We train the neural network 10,000 times, each time adopting a different random seed and varying the hyperparameters of the network as discussed below. This Monte Carlo approach allows us to obtain probability distribution functions (PDFs) of the target variables, where the resulting dispersion reflects the uncertainties of the neural network.

We follow the standard practice of training the network on the scaled feature variables, where across the $N_{\rm sat}$ data points we use a standard scaler to subtract the mean and divide by the standard deviation of each variable. The training set is randomly split into a training and a testing subset. For each Monte Carlo realisation of the neural network, we set the fraction of the training data that is used for testing the network by randomly drawing $f_{\rm test}$ from a flat distribution between 0.2 and 0.3. The network is compiled using the {\tt Adam} optimizer with the default hyperparameters within \keras and a mean squared error loss function. The neural network is fitted to the training data for a maximum of 50 epochs, but we use an early stopping monitor with a {\tt patience} value of 3. This means that the fitting loop is stopped when the validation score of the neural network does not improve for three successive epochs, so that in practice 50 fitting epochs are never reached. The model with the highest validation score is saved as a checkpoint and used when applying the model. The validation score is calculated by separating off a validation subset that is a fraction $f_{\rm val}$ of the training subset, where we randomly draw $f_{\rm val}$ from a flat distribution between 0.1 and 0.2 for each Monte Carlo realisation of the neural network. This means that the validation subset consists of $f_{\rm val}(1-f_{\rm test})N_{\rm sat}$ accretion events. We then test the neural network by using it to predict the target variables for the test subset and comparing them to the true values. The above procedure is repeated for each of the 10,000 Monte Carlo realisations. Throughout this paper, uncertainties on quoted numbers reflect the 16th and 84th percentiles of PDFs resulting (or propagated) from these 10,000 realisations. By drawing $f_{\rm test}$ and $f_{\rm val}$ from the aforementioned flat distributions for each Monte Carlo realisation, the uncertainties on the target variables account for the effects of changing these hyperparameters. Finally, we repeat the entire process when omitting certain subsets of the feature variables, i.e.\ when omitting (1) GC age information, (2) GC metallicity information, (3) GC age and metallicity information, and (4) GC orbital information. The goal of carrying out these additional experiments is to identify which of the feature variables are the best predictors of each of the target variables and should thus always be included in observational applications of these neural networks.

\autoref{fig:f1} shows the results of the above experiments.\footnote{Throughout this paper, PDFs reflect the 10,000 Monte Carlo realisations of the neural network. All PDFs are smoothened using the default kernel density estimator in the {\sc Seaborn} Python package \citep{waskom20}.} When including all feature variables, the neural network achieves a satisfactory validation score of $0.89^{+0.06}_{-0.06}$. This matches the typical training score to within the uncertainties, indicating that the model is not underfitting or overfitting. Omitting any of the feature variables results in lower validation scores, even if their medians are generally consistent with the validation score for all feature variables to within the scatter. The neural network only performs significantly worse when omitting the GC age and metallicity information, with a validation score of $0.76^{+0.07}_{-0.07}$. This shows that the GC age-metallicity distribution encodes crucial information on the stellar masses and accretion redshifts of the Milky Way's satellite progenitor galaxies. This lends further support to the rich body of literature identifying GC age-metallicity space as an important tracer of galaxy assembly \citep[e.g.][]{forbes10,leaman13,choksi18,kruijssen19e}.

We assess the precision with which the neural network predicts the stellar masses and accretion redshifts of the satellite progenitors by calculating the logarithmic standard deviation around the one-to-one relation for the test subset in each Monte Carlo realisation. The resulting scatter on the stellar mass is $\sigma[\log_{10}(M_{\rm \star,pred}/M_{\rm \star,true})]=0.41^{+0.05}_{-0.04}$ when including all feature variables. This precision enables making quantitative predictions for the satellite progenitor masses. As before, omitting the age-metallicity information leads to a significantly worse precision, with $\sigma[\log_{10}(M_{\rm \star,pred}/M_{\rm \star,true})]=0.62^{+0.06}_{-0.06}$. Interestingly, this is mostly driven by the GC metallicities, as omitting these leads to a scatter of $\sigma[\log_{10}(M_{\rm \star,pred}/M_{\rm \star,true})]=0.50^{+0.07}_{-0.06}$, whereas omitting only the GC age information barely affects the scatter around the one-to-one relation for true versus predicted satellite progenitor mass. We conclude that the GC metallicities, in combination with either GC ages or orbital information, is critical for constraining the stellar masses of the accreted galaxies.

For the accretion redshifts, we achieve a precision of $\sigma[\log_{10}(1+z_{\rm acc,pred}/1+z_{\rm acc,true})]=0.13^{+0.01}_{-0.01}$, or about 30~per~cent on $1+z_{\rm acc}$. Accurate predictions for the satellite accretion redshift require nearly all information to be used, with the exception of the GC metallicities. Omitting any other feature variable (GC ages or GC orbital information) leads to significantly worse constraints on the accretion redshift, with $\sigma[\log_{10}(1+z_{\rm acc,pred}/1+z_{\rm acc,true})]>0.16$. This shows that the accretion redshift is the most challenging of the two target variables to constrain. The reason it depends on GC ages is obvious -- satellites that were accreted early must have older GCs. The strong dependence on the orbital information is a bit more subtle, but also works as expected. Satellites that were accreted early deposit their GCs at small (apocentre) radii and these GCs end up with a large spread in eccentricities by $z=0$ \citep{pfeffer20}.

In summary, the inclusion of the complete GC age-metallicity information guarantees the most precise model predictions in general. The orbital information plays an important role in helping constrain the accretion redshift. In the following sections, we apply the model to the GC population of the Milky Way, for which all eight feature variables are known.

\section{Reconstructing the progenitor satellite population of the Milky Way} \label{sec:prog}

\subsection{Definition of the observational sample} \label{sec:obs}
\begin{figure*}
\includegraphics[width=\hsize]{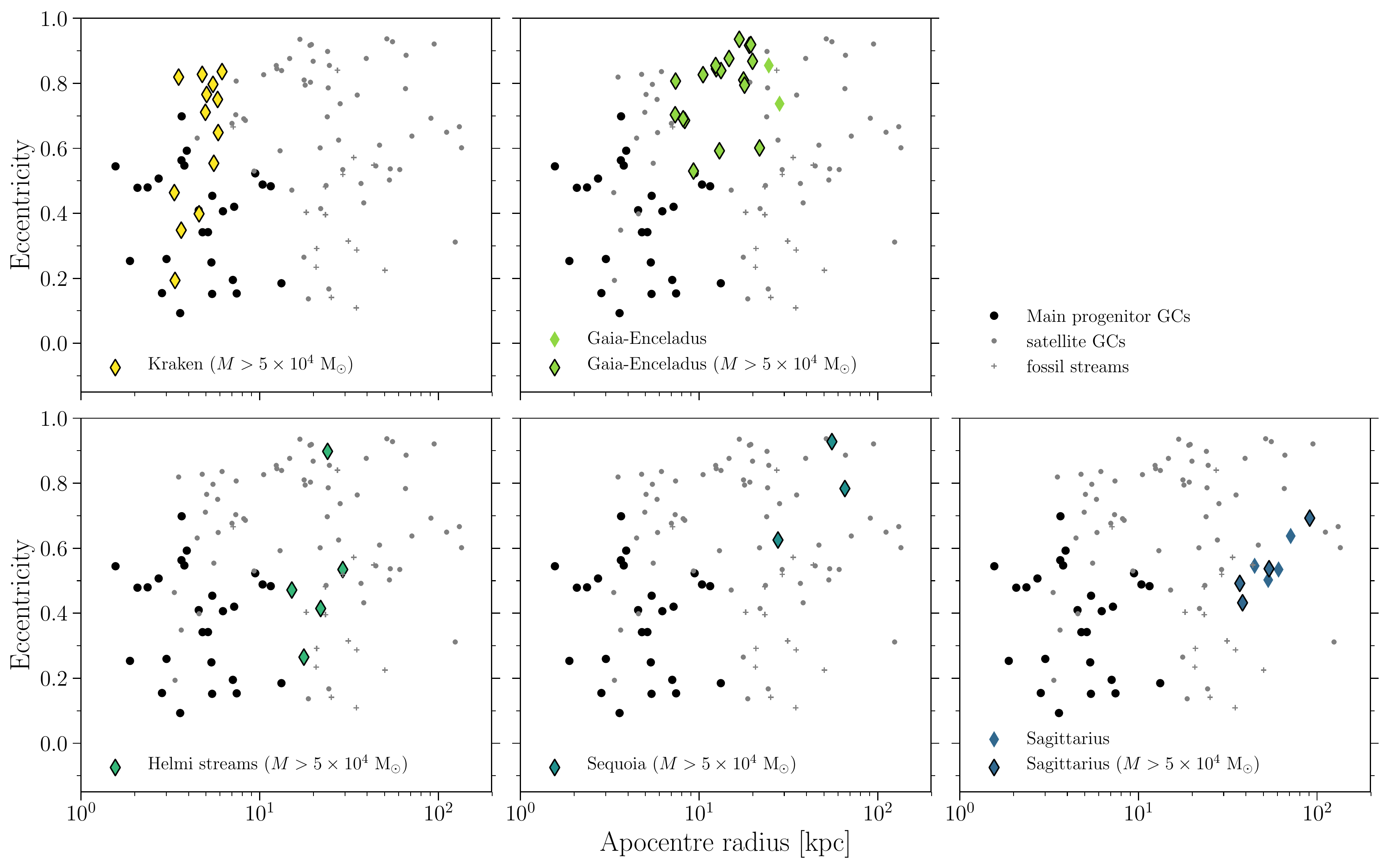}%
\caption{
\label{fig:f2}
Orbital properties of the Galactic GC population, expressed in terms of their apocentre radii and eccentricities. In all panels, we distinguish GCs that formed in the Main progenitor and those that formed in satellites (largely following \citealt{massari19}, with changes as discussed in the text), as well as fossil streams that are likely relics of GCs \citep[taken from][]{bonaca20}, which we do not associate to any specific accretion event here. In each of the panels, the unambiguous members of each satellite progenitor are highlighted, demonstrating that GCs associated with different satellite progenitors have different orbital properties.}
\end{figure*}
\citet{massari19} combine the orbital information of the Galactic GC system with their age-metallicity distribution to identify subsets of GCs that likely originated in a common progenitor. By dividing the GC population in this way, they identify five groups of GCs that plausibly have an extragalactic origin. Four of these they associate with known accretion events, i.e.\ the {\it Gaia}-Enceladus-Sausage event \citep{belokurov18,helmi18,myeong18}, the progenitor of the \citet{helmi99} streams, the Sequoia accretion event \citep{myeong19}, and Sagittarius \citep{ibata94}. In addition, they identify a group of GCs at low energies (i.e.\ at high binding energies), which is similar in number to the GCs associated with the \ges event. While \citet{massari19} are unable to draw firm conclusions regarding the origin of these low-energy GCs, we demonstrate below that this group represents the Kraken accretion event predicted by \citet{kruijssen19e}. After identifying these five groups, \citet{massari19} combine the remaining GCs with known ages, metallicities, and orbits in a `high-energy' group, which are distributed across parameter space and are unlikely to have originated in a common progenitor. Instead, they could represent an ensemble of low-mass accretion events that each contributed one or two GCs. We therefore omit the high-energy group from our further analysis.

We use the GC ages and metallicities from the compilation of \citet[who combined literature measurements from \citealt{forbes10}, \citealt{dotter10,dotter11}, and \citealt{vandenberg13}]{kruijssen19e} and the orbital properties from \citet{baumgardt19}. For the five satellite progenitors, we adopt the GC membership selection from \citet{massari19}, with a small number of changes. Firstly, they consider Pal~1 to have formed in-situ (i.e.\ in the `Main progenitor'). However, given its position in age-metallicity space, it unambiguously belongs to the satellite branch, having a relatively low metallicity of $\feh=-0.7$ at an age of only 7.3~Gyr. In order to have formed within the Main progenitor, Pal~1 would have needed to have had roughly solar metallicity \citep{haywood13}. Because it is not known to which satellite progenitor Pal~1 should be attributed, we omit if from our analysis. Conversely, \citet{massari19} associate NGC6441 and E3 with the low-energy group and (possibly) the progenitor of the Helmi streams, respectively. However, based on their high metallicities ($\feh=-0.6$ and $-0.83$) at old ages ($\tau=11.3\pm0.9$~Gyr and $\tau=12.8\pm1.4$~Gyr) these GCs must have formed in the Main progenitor. We therefore classify these GCs as ambiguous, possibly being members of the low-energy group or the Main progenitor, and the Helmi streams or the Main progenitor, respectively. In practice, this means that we exclude E3 from our analysis (also because of its low mass), and consider versions of the low-energy group both including and excluding NGC6441. Finally, \citet{horta20} argue that NGC6121, which \citet{massari19} associates with the low-energy group, has an in-situ origin. Based on the age and metallicity of NGC6121 ($\tau=12.2\pm0.5$~Gyr and $\feh=-1.14$), we consider it too metal-poor to have formed in-situ and follow the choice of \citet{massari19} to associate it with the low-energy group, even if we acknowledge that this is an edge case. The results presented in this paper are unaffected by this choice.

\begin{figure*}
\includegraphics[width=\hsize]{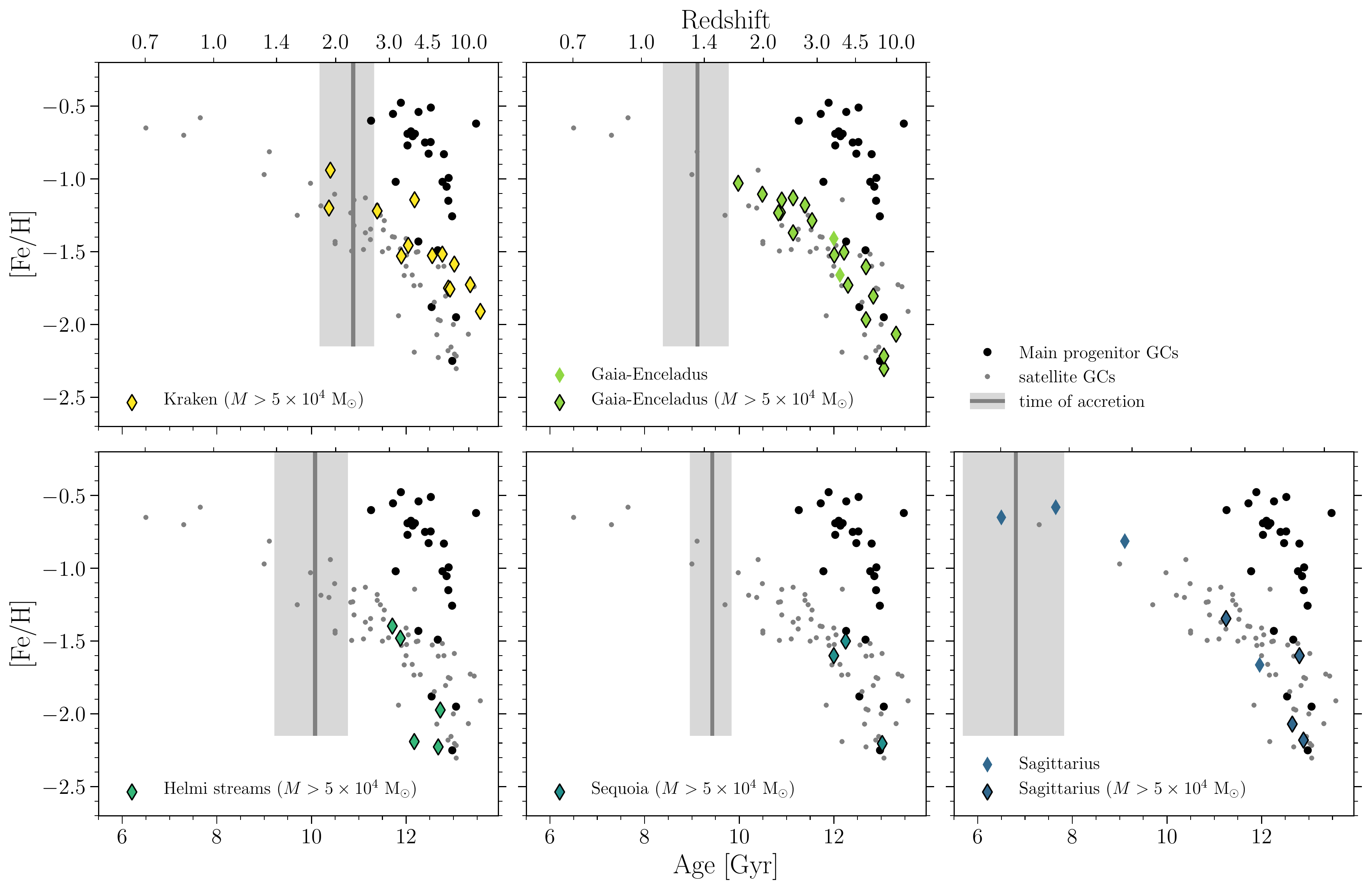}%
\caption{
\label{fig:f3}
Age-metallicity distribution of the Galactic GC population. In all panels, we distinguish GCs that formed in the Main progenitor and those that formed in satellites (largely following \citealt{massari19}, with changes as discussed in the text). In each of the panels, the unambiguous members of each satellite progenitor are highlighted, demonstrating that GCs associated with different satellite progenitors follow different tracks in age-metallicity space. Each panel also includes a vertical line indicating the time of accretion inferred in Section~\ref{sec:app}, with the shaded band representing the $1\sigma$ uncertainty.}
\end{figure*}
Before proceeding, we note that not all memberships proposed by \citet{massari19} are unambiguous. Several GCs have dual associations (e.g.\ `low-energy/Sequoia'). We restrict our fiducial analysis to the GCs that are unambiguously associated with a single satellite progenitor. However, throughout this paper we systematically consider all possible membership permutations, i.e.\ all possible assignments of ambiguous GCs to the satellite progenitors that \citet{massari19} identify as possible hosts. The the goal of this procedure is to demonstrate how the uncertainties in GC membership propagate into uncertainties on the stellar masses and accretion redshifts of the satellite progenitors. From hereon, we also refer to the low-energy GCs of \citet{massari19} as Kraken GCs.

We show the distributions of Galactic GCs in the plane spanned by apocentre radius and eccentricity in \autoref{fig:f2} and in the plane spanned by age and metallicity in \autoref{fig:f3}, colour-coded by their memberships of each of the five satellite progenitors. In the colour coding, we only include the unambiguous memberships. In addition, here and in the subsequent analysis we include observed GCs with masses $M<5\times10^4~\msun$, contrary to the selection of the training set. This mass cut only needs to be applied to the simulations, because it compensates the underdisruption of GCs in \emosaics -- real-Universe GCs do not suffer from this problem, implying that a mass cut is not required. The inclusion of low-mass GCs has the additional benefit that the statistics of the observational samples improve somewhat.

In \autoref{fig:f2}, the five groups of GCs occupy distinct parts of apocentre-eccentricity space. The Kraken GCs occupy the smallest radii, but a considerably wider range of eccentricities than the other satellites, or even than the Main progenitor. This suggests that Kraken was massive and accreted early \citep{pfeffer20}. The \ges GCs occupy intermediate radii and high eccentricities, whereas the GCs associated with the Helmi streams orbit at intermediate radii and (mostly) intermediate eccentricities. Finally, the Sequoia and Sagittarius GCs have large apocentre radii, with high and intermediate eccentricities, respectively. These differences are suggestive of differences in origin -- the orbital characteristics of each group of GCs reflect the orbital properties of the accretion events, which in turn trace the masses and accretion redshifts of the satellite progenitors.

For reference, \autoref{fig:f2} also includes the orbital properties of several fossil stellar streams in the Galactic halo, taken from \citet{bonaca20}, which plausibly originated from disrupted GCs. The reasonable correspondence between the apocentre-eccentricity distribution of these streams and that of the GCs associated with each of the five progenitor satellites implies that several of these streams may be associated with the same progenitors. In general, the streams seem to be relics of GCs with an ex-situ origin, as there is little correspondence with the orbital properties of GCs that formed in the Main progenitor. The orbital eccentricities of most streams do not reach values as high as those of \ges and Sequoia, but instead occupy the low-to-intermediate eccentricity range. We find one stream (Fimbulthul) that may possibly be a relic of a GC that formed in Kraken (here selected using $\ra\la7~\kpc$). Two streams (again Fimbulthul, but also Slidr) may be associated with \ges, given their high orbital eccentricities ($\ecc\ga0.55$) and intermediate radii ($7\la\ra/\kpc\la30$). A further seven streams (Fj\"{o}rm, Gj\"{o}ll, Indus, Phoenix, Tucana~III, Turranburra, Ylgr) orbit at radii ($15\la\ra/\kpc\la30$) and eccentricities ($0.2\la\ecc\la0.55$) similar to those of the GCs associated with the Helmi streams, suggesting that these seven streams could have originated from GCs that were once part of the \citet{helmi99} satellite progenitor. There are no streams unambiguously associated with the Sequoia GCs, except possibly Gj\"{o}ll and Turranburra (this depends critically on the membership of IC~4499). Finally, two streams (Leiptr and Turranburra) have apocentre radii $\ra\ga30~\kpc$ and eccentricities $0.4\la\ecc\la0.7$, suggesting that they could be disrupted GCs that were brought in by Sagittarius. We refer to \citet{bonaca20} for more details on the properties of the fossil streams and conclude this brief discussion by emphasising that the selection performed here is generous -- the inclusion of additional selection criteria may further trim the sample of streams associated with the satellite progenitors.

In \autoref{fig:f3}, we show the age-metallicity distribution of the GC sample. As for \autoref{fig:f2}, it is immediately clear that the GCs associated with the five satellite progenitors occupy a different part of the plane than those that formed in-situ within the Main progenitor. The figure also shows that there are quantitative differences between the individual satellites. At any given age, the Kraken GCs are the most metal rich, which suggests that Kraken was the most massive satellite at any given moment in time prior to its accretion onto the Milky Way (in terms of both its stellar and halo mass).\footnote{Despite possibly being the most massive at any time prior to its accretion, Kraken need not be the most massive satellite that the Milky Way ever accreted, if it was accreted significantly earlier than the other satellites. This would truncate its growth, whereas satellites that were accreted later would be able to continue growing their masses. See Section~\ref{sec:app} for details.} The second most massive satellite at any given time is \ges, with the progenitor of the Helmi streams, Sequoia, and Sagittarius, likely having had lower masses at any time, based on the fact that the metallicities of their GCs are generally lower and their GC populations are less numerous at young ages ($\tau<11~\gyr$). Finally, we see no GCs with ages younger than the accretion redshifts inferred in Section~\ref{sec:app}, which is an important consistency check. We do note that the most massive satellites (Kraken, \ges, and Sagittarius, see Section~\ref{sec:ms}) are able to form GCs all the way up to their time of accretion, whereas the lower-mass satellites (the progenitor of the Helmi streams and Sequoia, also see Section~\ref{sec:ms}) have GC formation truncated earlier. This may happen because the low-mass satellites are getting disrupted more rapidly after they enter the Galactic halo, whereas the massive satellites survive longer.

\subsection{Application of the neural network} \label{sec:app}
We take the neural network described in Section~\ref{sec:nn} and apply it to the GC populations associated with each of the satellite progenitors to obtain their stellar masses and accretion redshifts. We list the adopted memberships in Table~\ref{tab:member}. As discussed above, some of the memberships are ambiguous. Throughout the majority of the following discussion, we consider all possible permutations that at least include the unambiguous members of each satellite progenitor. For progenitors that are listed $\{1,2,3,4\}$ times in Table~\ref{tab:member}, this means we need to consider $\{1,2,4,8\}$ possible permutations, for a total of 23 permutations across all five satellite progenitors.

For each Monte Carlo realisation of the neural network and for each membership permutation, we subject the eight feature variables (medians and IQRs of the GC age, metallicity, apocentre radius, and eccentricity) to the same scaler transform as we did for the training set in Section~\ref{sec:nndetails}, i.e.\ we subtract the mean of the training set and divide by the standard deviation of the training set. This transforms the GC properties to the same coordinate space used to train the network. We then use the neural network to predict the two target variables (satellite progenitor stellar mass and its accretion redshift). This results in 10,000 predictions for each target variable and for each GC membership permutation.
\begin{table*}
  \caption{GC membership adopted in this work. For each satellite progenitor, we consider all possible membership permutations throughout our analysis. The `Abbreviation' column lists the shorthand used to refer to these subsets of GCs in figure legends.}
\label{tab:member}
  \begin{tabular*}{\textwidth}{l @{\extracolsep{\fill}} l l}
   \hline
   Possible progenitors & Abbreviation & GCs \\
   \hline
   \vspace{0.3mm}Kraken & -- & \makecell[l]{NGC5946, NGC5986, NGC6093, NGC6121, NGC6144, NGC6254, NGC6273, \\ NGC6287, NGC6541, NGC6544, NGC6681, NGC6712, NGC6809} \\
   \vspace{0.3mm}Kraken/Main progenitor & Kraken/Main & NGC6441 \\ 
   \vspace{0.3mm}Kraken/Sequoia & Kraken/Seq & NGC6535 \\ 
   \vspace{0.3mm}\ges & G-E & \makecell[l]{NGC288, NGC362, NGC1261, NGC1851, NGC1904, NGC2298, NGC2808, \\ NGC4147, NGC4833, NGC5286, NGC5897, NGC6205, NGC6235, NGC6284, \\ NGC6341, NGC6779, NGC6864, NGC7089, NGC7099, NGC7492} \\ 
   \vspace{0.3mm}\ges/Sequoia & G-E/Seq & NGC5139 \\ 
   \vspace{0.3mm}Helmi streams & H99 & NGC4590, NGC5024, NGC5053, NGC5272, NGC6981 \\ 
   \vspace{0.3mm}Helmi streams/\ges & H99/G-E & NGC5634, NGC5904 \\ 
   \vspace{0.3mm}Sequoia & Seq & NGC5466, NGC7006, IC4499 \\ 
   \vspace{0.3mm}Sequoia/\ges & Seq/G-E & NGC3201, NGC6101 \\ 
   Sagittarius & -- & NGC2419, NGC5824, NGC6715, Pal 12, Terzan 7, Terzan 8, Arp 2, Whiting 1 \\ 
   \hline
  \end{tabular*} 
\end{table*}

\subsubsection{Stellar masses at the time of accretion} \label{sec:ms}
\begin{figure*}
\includegraphics[width=\hsize]{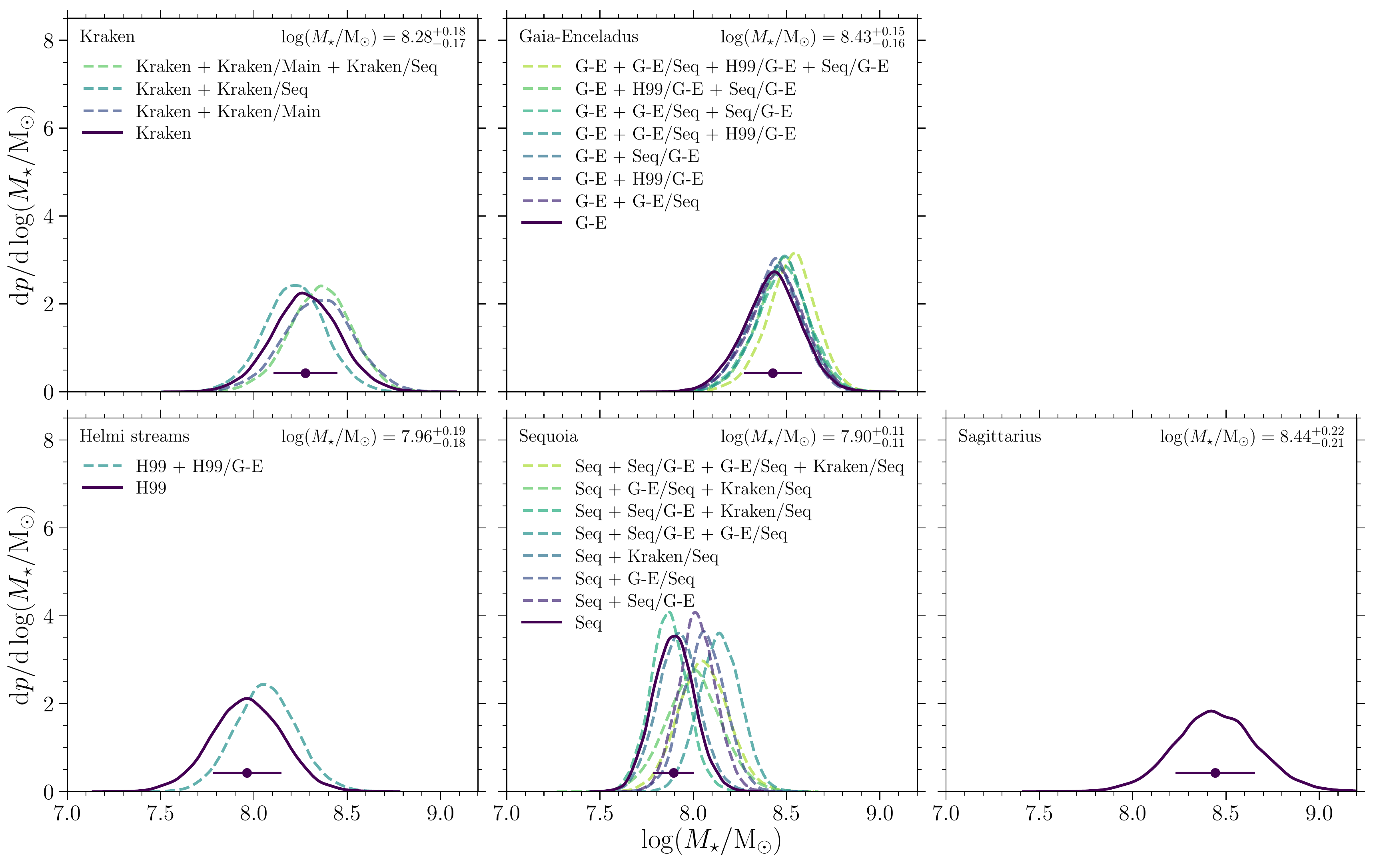}%
\caption{
\label{fig:f4}
PDFs of the predicted stellar masses of the satellite progenitors, inferred by applying a neural network trained on the \emosaics simulations to their GC populations. Each PDF shows the distribution across all 10,000 Monte Carlo realisations of the neural network, applied to different GC membership permutations as indicated by the legend (also see Table~\ref{tab:member}). In each panel, the data point with error bar and the annotation in the top right indicate the median and 16th-to-84th percentiles for the experiment using the unambiguous member GCs (corresponding to the solid line in that panel). This figure shows that the five satellite progenitors span a relatively narrow stellar mass range, of $\ms=(0.6{-}4.6)\times10^8~\msun$.}
\end{figure*}

\autoref{fig:f4} shows the PDFs of the stellar mass of the satellite progenitors at the time of accretion (see Section~\ref{sec:sims} for its definition). The satellite progenitors span a relatively narrow range of stellar masses at the time of accretion, of a factor of 3--4. As expected, Kraken and \ges are the main accretion events, with stellar masses of $\ms=1.9^{+1.0}_{-0.6}\times10^8~\msun$ and $\ms=2.7^{+1.1}_{-0.8}\times10^8~\msun$, respectively. These masses exceed those of the progenitor of the Helmi streams and Sequoia, which had stellar masses of $\ms=0.9^{+0.5}_{-0.3}\times10^8~\msun$ and $\ms=0.8^{+0.2}_{-0.2}\times10^8~\msun$, respectively. Somewhat surprisingly, Sagittarius is predicted to have had a mass of $\ms=2.8^{+1.8}_{-1.1}\times10^8~\msun$, similar to Kraken and \ges when they merged, despite the Sagittarius GCs having lower metallicities when considering the same GC age interval. This high mass is enabled by the late time of its accretion, giving it more time to grow its stellar mass than Kraken and \ges had before they got cannibalised (see Section~\ref{sec:zacc}).

The stellar masses that we infer for each of the satellite progenitors are only weakly affected by the GC membership permutation. The total spread of the median falls within the uncertainties on the prediction for all progenitors except Sequoia, for which the inclusion of ambiguous GCs can increase its mass by up to 0.2~dex. This is not surprising, because a large fraction (40~per~cent) of the GCs potentially associated with Sequoia is ambiguous. For all other progenitors, the impact of the membership ambiguity is smaller and usually of the order 0.1~dex.

Relative to other measurements in the literature, we find a lower mass for \ges than \citet{helmi18}, who provide a very rough estimate of $\ms\sim6\times10^8~\msun$ from the star formation rate ($\sim0.3~\msun~\yr^{-1}$) and duration ($\sim2~\gyr$) necessary to reproduce the $\alpha$-poor stellar population \citep{fernandezalvar18}. The back-of-the-envelope nature of this estimate makes it quite uncertain and, depending on the membership of the $\alpha$-poor stellar population, it may represent an upper limit. We therefore consider our predicted stellar mass broadly compatible with the estimate of \citet{helmi18}, but point out that the prediction made here is likely to be more accurate, as well as more meaningful thanks to the inclusion of error bars on the predictions. In \citet{kruijssen19e}, we estimated that both Kraken and \ges had masses as high as $10^9~\msun$ at the time of accretion, based on the age-metallicity-mass distribution of central galaxies in the \eagle simulation. In the present paper, we place them at a lower mass, because the satellites that are accreted onto Milky Way-mass centrals have lower masses than field dwarf galaxies, as their star formation may be halted soon after falling into the halo. The neural network trained here automatically accounts for this bias, which was left unaccounted for by \citet{kruijssen19e}.

The stellar masses inferred for the final three satellite progenitors also agree with previous literature results. \citet{koppelman19} estimate that the progenitor of the Helmi streams had a mass of $\ms\sim10^8~\msun$ when it accreted, which is entirely consistent with our prediction. Likewise, \citet{myeong19} estimate that Sequoia had a stellar mass at the time of accretion of $\ms=1.7\times10^8~\msun$, with a factor-of-few uncertainty, again consistent with the mass reported above. Finally, \citet{niedersteostholt10,niedersteostholt12} find that the stellar mass of Sagittarius at the time of accretion was $\ms=(2.0{-}2.9)\times10^8~\msun$ (assuming a mass-to-light ratio of $M/L=2~\msun~{\rm L}_\odot^{-1}$), compatible with our prediction. Sagittarius is by far the best-studied accretion event out of the five considered here. The fact that the neural network's prediction for the mass of this galaxy is in such good agreement with independent mass estimates from the literature adds credence to the network's predictions for the masses of the other satellite progenitors (as well as their accretion redshifts, see Section~\ref{sec:zacc}).

\citet{forbes20} estimate the stellar masses of the satellite progenitors considered here by using the total number of GCs (including ambiguous members) as a probe of the halo mass and converting it to a stellar mass by adopting a stellar mass--halo mass relation. This makes the strong assumption that the relation between the number of GCs and the halo mass at $z=0$ does not evolve with redshift. If the number of GCs per unit halo mass is higher at high redshift \citep[as suggested by e.g.][Bastian et al.\ in prep.]{kruijssen15b,choksi19,elbadry19}, this assumption overestimates the galaxy mass. Likewise, the inclusion of ambiguous GCs as members also maximises the galaxy mass. For these reasons, the resulting estimates of the stellar masses represent (quite uncertain) upper limits, because they effectively represent projected masses at $z=0$ rather than at the time of accretion. For Kraken, \ges, the progenitor of the Helmi streams, Sequoia, and Sagittarius, \citet{forbes20} estimates $\log{(\ms/\msun)}=\{8.7,8.9,7.9,7.9,7.9\}$. Out of these, the masses of the progenitor of the Helmi streams and Sequoia are consistent with the masses derived here. The masses of Kraken and \ges are larger, most likely due to the biases described above. The mass of Sagittarius is lower than both the result of \citet{niedersteostholt12} and the mass derived in this paper.

\begin{figure*}
\includegraphics[width=\hsize]{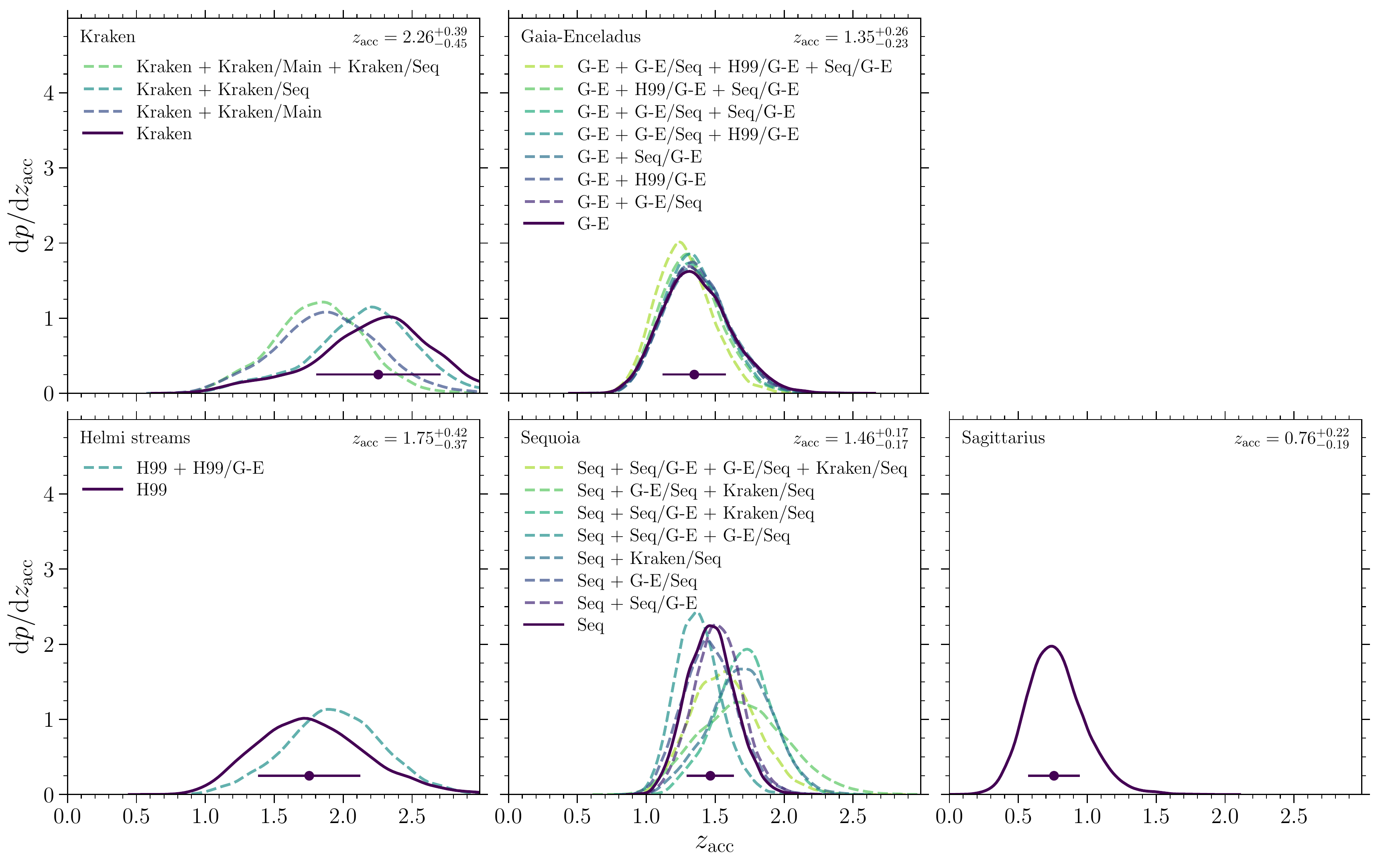}%
\caption{
\label{fig:f5}
PDFs of the predicted accretion redshifts of the satellite progenitors, inferred by applying a neural network trained on the \emosaics simulations to their GC populations. Each PDF shows the distribution across all 10,000 Monte Carlo realisations of the neural network, applied to different GC membership permutations as indicated by the legend (also see Table~\ref{tab:member}). In each panel, the data point with error bar and the annotation in the top right indicate the median and 16th-to-84th percentiles for the experiment using the unambiguous member GCs (corresponding to the solid line in that panel). This figure shows that the five satellite progenitors accreted over a wide redshift range of $\zacc=0.57{-}2.65$, corresponding to lookback times of $\tacc=5.7{-}11.3~\gyr$.}
\end{figure*}
We conclude the discussion of \autoref{fig:f4} by pointing out that the quoted error bars on the predictions reflect random uncertainties. A comparison to the standard deviation shown in the middle panel of \autoref{fig:f1} (which is larger than the typical error bar) suggests the existence of an additional systematic uncertainty that may affect all stellar mass measurements by up to 0.3~dex.

\subsubsection{Accretion redshifts} \label{sec:zacc}
\autoref{fig:f5} shows the PDFs of the accretion redshifts of the satellite progenitors, which is defined in \emosaics as the moment at which we can no longer detect a gravitationally bound subhalo (see Section~\ref{sec:sims}). As is immediately obvious from the figure, the satellite progenitors span a wide range of accretion redshifts. Chronologically, Kraken was the first galaxy to be accreted ($\zacc=2.26^{+0.39}_{-0.45}$ or $\tacc=10.9^{+0.4}_{-0.7}~\gyr$), followed by the progenitor of the Helmi streams ($\zacc=1.75^{+0.42}_{-0.37}$ or $\tacc=10.1^{+0.7}_{-0.9}~\gyr$). Sequoia ($\zacc=1.46^{+0.17}_{-0.17}$ or $\tacc=9.4^{+0.4}_{-0.5}~\gyr$) and \ges ($\zacc=1.35^{+0.26}_{-0.23}$ or $\tacc=9.1^{+0.7}_{-0.7}~\gyr$) accreted at approximately the same time, but still well before Sagittarius ($\zacc=0.76^{+0.22}_{-0.19}$ or $\tacc=6.8^{+1.1}_{-1.1}~\gyr$), which is the final accretion event considered here. For all five satellite progenitors the lookback time of accretion is consistent with (or larger than) the age of the youngest associated GC, which provides an important consistency check (see \autoref{fig:f3}).

The large variety of accretion redshifts explains an apparent inconsistency that appeared above. In the discussion of \autoref{fig:f3}, we previously inferred the mass-ranked order of satellites at a given age from their metallicity offsets and suggested that Kraken was the most massive satellite, whereas Sagittarius was one of the lowest-mass ones. As discussed above, this differs from their inferred mass-ranked order at the time of their accretion (see \autoref{fig:f4}). We now see that this difference arises, because satellites that were accreted early had their mass growth cut short, whereas those that were accreted late continued to grow long after the other, initially more massive satellites were disrupted.

For a subset of the satellite progenitors, the GC membership permutation influences the accretion redshift more strongly than it affects the stellar masses (see the discussion of \autoref{fig:f4}). When expanding the sample of unambiguous Kraken GCs with NGC6441, which most likely formed in-situ in the Main progenitor (see Section~\ref{sec:obs}), the accretion redshift decreases considerably, to $\zacc\approx1.8$. However, this is extremely unlikely to be accurate given the high metallicity ($\feh=-0.6$) and old age ($\tau=11.3\pm0.9~\gyr$) of NGC6441. When omitting NGC6441 from the ex-situ sample, the GC membership permutation has no significant influence on the accretion redshift of Kraken. Likewise, the accretion redshifts of \ges and the progenitor of the Helmi streams are not significantly affected by the GC membership selection, as the shifts of the median fall well within the quoted uncertainties. As for its stellar mass, the accretion redshift of Sequoia changes for different GC membership permutations. Specifically, when attributing the ambiguous GC NGC6535 to Sequoia, it systematically has a higher accretion redshift of $\zacc\approx1.7$. While the position of NGC6535 in age-metallicity space ($\feh=-1.73$ and $\tau=12.2\pm0.6~\gyr$) does not allow distinguishing between Kraken and Sequoia, its orbital properties ($\ra\approx4.5~\kpc$ and $\ecc\approx0.63$) clearly place it in a part of the orbital parameter space that is not covered by any of the Sequoia GCs (which occupy $\ra>25~\kpc$ in the selection of \citealt{massari19}) and is rather consistent with the Kraken GCs (which have $\ra<7~\kpc$).\footnote{We note that \citet{myeong19} associate additional GCs with Sequoia that also have apocentres $\ra<10~\kpc$ (NGC5139 and NGC6388). The first of these is included in our other permutations for Sequoia (see Table~\ref{tab:member}), implying that its inclusion or omission does not affect our results, whereas we omit NGC6388 altogether, because its age ($\tau=12.0\pm1.0~\gyr$) and metallicity ($\feh=-0.77$) imply that it likely formed in-situ.} When omitting NGC6535 from the Sequoia membership permutations, the accretion redshift is always consistent with the value reported for the unambiguous GC membership assignment.

Relative to other measurements in the literature, our predicted accretion redshifts largely satisfy the constraints obtained through independent methods. \citet{helmi18} estimate that \ges was accreted $\sim10~\gyr$ ago, whereas \citet{belokurov18} estimate a range of $8{-}11~\gyr$ ago, and \citet{mackereth19} provide an upper limit on the accretion redshift of $\zacc<1.5$ (or $\tacc<9.5~\gyr$). All three of these constraints are consistent with our predicted range of $\tacc=9.1^{+0.7}_{-0.7}~\gyr$. The same applies for Sequoia, which has been proposed to have been accreted $9{-}11~\gyr$ ago \citep{myeong19}, consistently with our estimate of $\tacc=9.4^{+0.4}_{-0.5}~\gyr$. By contrast, \citet{koppelman19} propose that the progenitor of the Helmi streams was accreted $5{-}8~\gyr$ ago, whereas our analysis predicts $\tacc=10.1^{+0.7}_{-0.9}~\gyr$. As discussed above, extending the GC membership leads to even later accretion redshifts. The estimate of \citet{koppelman19} is based on collisionless $N$-body simulations of the Helmi streams, obtaining a best kinematic match for accretion times of $5{-}8~\gyr$. However, they do find that the stellar age range of the Helmi streams is $\tau=11{-}13~\gyr$. The agreement of the lower bound of the age range with our predicted accretion time suggests that star formation in the progenitor of the Helmi streams may have been truncated by its tidal disruption in the Galactic halo. If this interpretation is correct, then it remains an important open question how the dynamical constraints from \citet{koppelman19} can be reconciled with this picture. There are several possible explanations. The upper limit on the accretion time reported by \citet{koppelman19} of $8~\gyr$ reflects the maximum of the range of accretion times considered in their dynamical models, indicating that earlier accretion times may be dynamically possible, but have not been explored. Additionally, the dynamical models assume that the orbits do not evolve in time and neglect dynamical friction. In view of these considerations, a plausible solution would be to extend the orbital parameter space surveyed by \citet{koppelman19} to look for a kinematic match that also satisfies the prediction of our model.

For Sagittarius, the simulations of \citet{law10} and \citet{niedersteostholt12} suggest that it has been undergoing intense tidal disruption for the past $4{-}7~\gyr$. Which moment in this interval corresponds to our time of accretion (as obtained from the \emosaics simulations) depends quite sensitively on its definition. The large time interval over which the disruption of Sagittarius has been taking place greatly complicates this interpretation. At face value, the time over which Sagittarius has been strongly disrupted is consistent with our prediction of $\tacc=6.8^{+1.1}_{-1.1}~\gyr$. Using the association of GCs with stellar streams from disrupted dwarf satellites in \emosaics, \citet{hughes19} find a relation between the age range of accreted GCs and the stellar mass of the satellite progenitor, where excursions from that relation are strongly correlated with the infall time, i.e.\ the time at which the progenitor crosses the virial radius of the halo. Applying these relations to Sagittarius, \citet{hughes19} predict that it entered the halo of the Milky Way $t_{\rm infall}=9.3\pm1.8~\gyr$ ago. This upper limit on the time of accretion is consistent with our prediction, as well as with the long time (at least $4{-}7~\gyr$) spent by Sagittarius in the Galactic halo.

Finally, we point out that the quoted error bars on the predictions reflect random uncertainties. A comparison to the standard deviation shown in the right-hand panel of \autoref{fig:f1} (which is larger than the typical error bar) suggests the existence of an additional systematic uncertainty that may affect all accretion redshift measurements by up to 0.25 points in redshift.

\subsubsection{Scaling relations between GC sub-populations and their host satellite progenitor masses} \label{sec:scaling}
There exist well-documented relations between the total number ($\ngc$) or mass ($\mgcs$) of GCs and the stellar ($\ms$) or halo virial ($\mh$) mass of the host galaxy \citep[e.g.][]{spitler09,durrell14,hudson14,harris17,forbes18b,burkert20}. Because these relations are close to linear, they are often expressed in terms of number ratios, such as the specific frequency normalised by stellar mass ($\tn\equiv\ngc/\ms$) or by halo mass ($\etan\equiv\ngc/\mh$), or in terms of the ratio between the total GC system mass and the halo virial mass ($\etam\equiv\mgcs/\mh$). It is an important question whether these relations are fundamental and were imprinted at the time of GC formation \citep{spitler09,boylankolchin17,harris17,burkert20}, or if they result from a combination of ongoing baryonic processes and further linearisation by hierarchical galaxy assembly \citep[Bastian et al.\ in prep.]{kruijssen15b,choksi18,vandokkum18b,elbadry19}. The key underlying question is how these metrics evolve with redshift. If they are imprinted at birth, they should exhibit little redshift evolution. However, if they are the outcome of gradual galaxy formation processes, then they should evolve with redshift. It is hard to avoid this interpretation -- the fact that GCs are disrupted and galaxies grow with time means that the above metrics ($\tn$, $\etan$, and $\etam$) should be higher at higher redshifts. Now that we have obtained the stellar masses, accretion redshifts, and number of GCs of the satellite progenitors, we can infer the above metrics at the times of satellite accretion and compare them to the relations observed across the $z=0$ galaxy population. In addition to shedding light on the nature of the above scaling relations, placing the satellite progenitors in the context of these scaling relations also serves as an important consistency check.

To calculate the halo virial mass at the time of accretion for each of the satellite progenitors, we use the semi-empirical determination of the average relation between the stellar mass and halo mass as a function of redshift from \citet{moster13}, which is constrained by the observed evolution of the galaxy stellar mass function. We use their analytical expression to predict the stellar mass as a function of halo mass and redshift, and then numerically invert it by cubic interpolation to calculate the halo mass over the halo mass range $\mh=10^{10}{-}10^{12}~\msun$ and the redshift range $z=0{-}5$. In addition, we obtain the number of GCs by simply counting the members (see Table~\ref{tab:member}), and the total mass of the GC system by adding up the individual (dynamical) GC masses from \citet{baumgardt18}. Strictly speaking, this provides a lower limit, because there is no guarantee that the GC membership list is complete. Likewise, we necessarily omit any effects of GC disruption in the Milky Way halo -- if we could assign any of the fossil streams shown in \autoref{fig:f2} to individual satellite progenitors, this would also increase the number of GCs and their total mass. None the less, it allows us to calculate (lower limits on) $\tn$, $\etan$, and $\etam$ at the time of accretion for each of the five satellite progenitors and their GC membership permutations. Uncertainties on the resulting numbers are obtained by calculating the 16th and 84th percentiles across all 10,000 Monte Carlo realisations of the neural network.

\begin{figure*}
\includegraphics[width=\hsize]{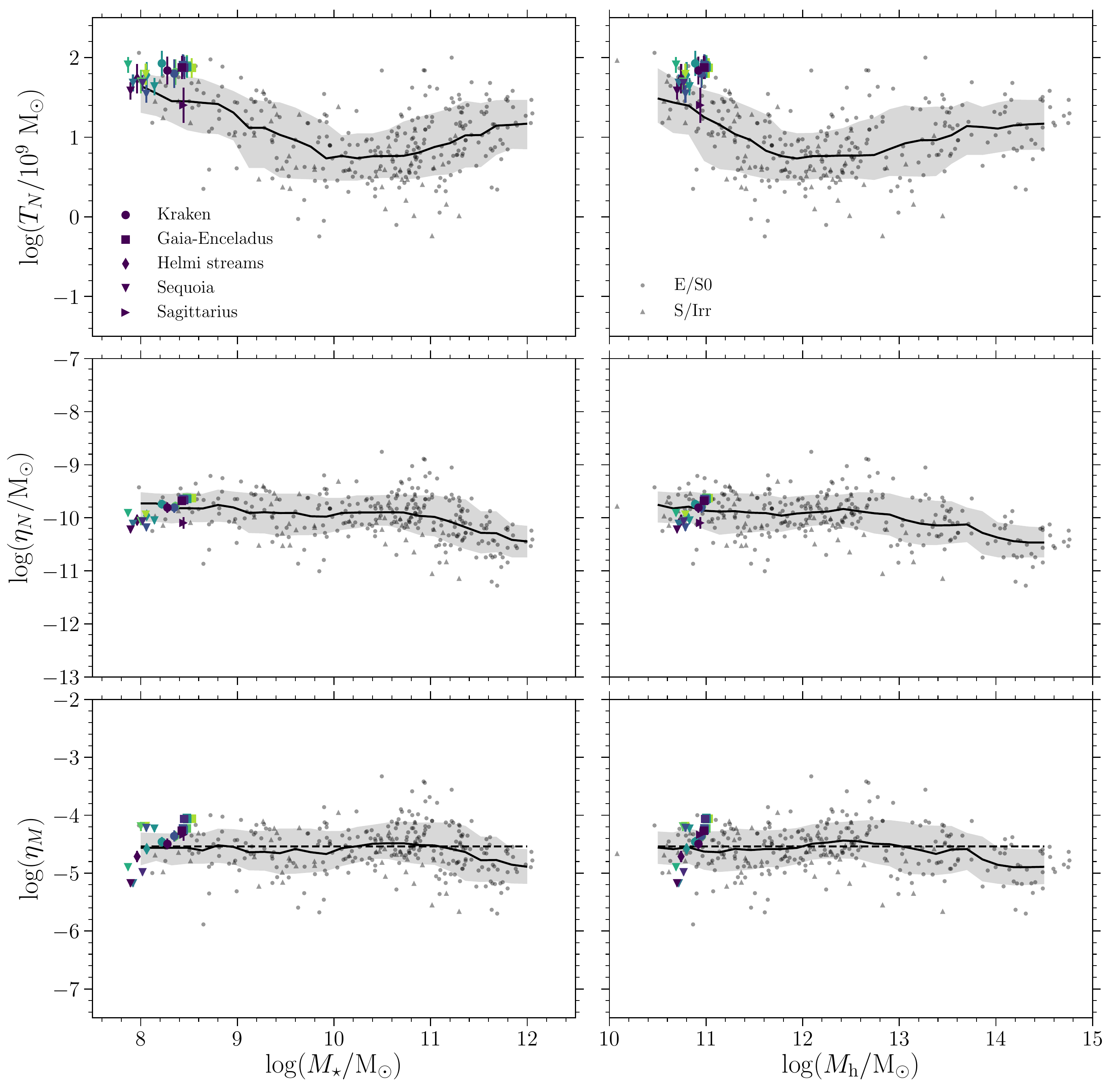}%
\caption{
\label{fig:f6}
GC specific frequency (top row, $\tn\equiv\ngc/\ms$), number of GCs per unit halo virial mass (middle row, $\etan\equiv\ngc/\mh$), and ratio between the GC system mass and halo mass (bottom row, $\etam\equiv\mgcs/\mh$), shown as a function of the galaxy stellar mass (left-hand column) and halo mass (right-hand column). The predictions for the satellite progenitors considered in this work are shown by the coloured symbols, with colours matching the GC membership permutations from \autoref{fig:f4} and~\ref{fig:f5}, and symbol shapes referring to the different progenitors as indicated by the legend. The observed galaxy population at $z=0$ \citep{harris17} is represented by the small grey symbols, showing early-type (E/S0) and late-type (S/Irr) galaxies. The solid line shows the running median across a 1~dex window, with 16th and 84 percentiles indicated by the grey-shaded band. In the bottom row, the horizontal dashed line indicates the roughly constant value of $\etam=2.9\times10^{-5}$ observed at $z=0$ \citep{harris17}.}
\end{figure*}
\autoref{fig:f6} shows how $\tn$, $\etan$, and $\etam$ vary as a function of $\ms$ and $\mh$, both for the satellite progenitors considered here and for the $z=0$ galaxy sample from \citet{harris17}, which includes 257 early-type galaxies (E/S0) and 46 late-type galaxies (S/Irr). In general, the satellites satisfy the scaling relations between the number or mass of GCs and the halo mass, exhibiting a similar scatter of $\etan$ and $\etam$ as the $z=0$ galaxy population. However, for the specific frequency ($\tn$), they fall above the observed relation, suggesting that the relation between $\ngc$ and $\ms$ (or $\mh$) evolved since the accretion redshifts of the satellite progenitors. Interestingly, the satellite that accreted the most recently (Sagittarius) is the most consistent with the observations, whereas the satellite that accreted the earliest (Kraken) shows the strongest excess, together with \ges, falling 0.5~dex above the observations at $z=0$. This is no firm evidence, but at least suggests that the stellar mass of isolated galaxies with halo masses around $\mh\sim10^{11}~\msun$ grew by a factor of $\sim3$ in stellar mass since $\zacc\sim1$. This is consistent with the abundance matching models of \citet{moster13}, who predict a stellar mass growth by a factor of $\sim5$ over the same time interval, and of \citet{behroozi13}, who predict a factor of $\sim4$. If true, this result implies that future observations of GC populations at $z>1$ should find higher specific frequencies.\footnote{This prediction is not affected directly by the details of the GC formation and disruption model in \emosaics, because we use the observed numbers of GCs rather than the numbers of GCs produced in the \emosaics simulations. Of course, the inferred satellite progenitor masses (i.e.\ the denominators of $\tn$, $\etan$, and $\etam$) do rely on the GC demographics from \emosaics, but the medians and IQRs of their ages, metallicities, apocentre radii, and eccentricities are not as strongly affected by the details of GC formation and disruption as their absolute numbers are.}

\begin{figure*}
\includegraphics[width=\hsize]{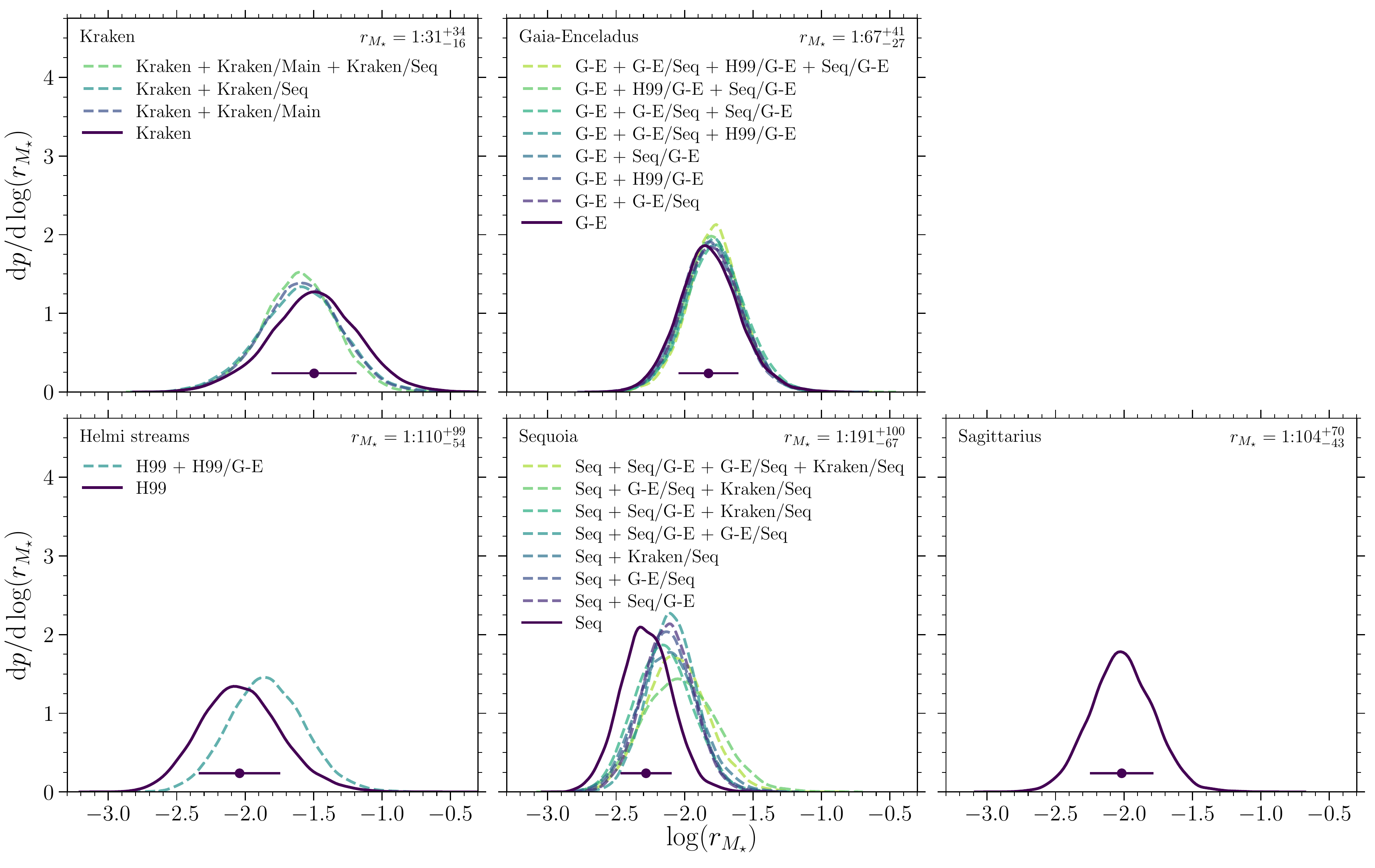}%
\caption{
\label{fig:f7}
PDFs of the predicted merger stellar mass ratios of the satellite progenitor accretion events onto the Milky Way, inferred by applying a neural network trained on the \emosaics simulations to their GC populations. Each PDF shows the distribution across all 10,000 Monte Carlo realisations of the neural network, applied to different GC membership permutations as indicated by the legend (also see Table~\ref{tab:member}). In each panel, the data point with error bar and the annotation in the top right indicate the median and 16th-to-84th percentiles for the experiment using the unambiguous member GCs (corresponding to the solid line in that panel). This figure shows that the five accretion events are all minor mergers, spanning a wide range of mass ratios $\rmass=1$:$(15{-}291)$. The accretion of Kraken likely represents the most major merger that the Milky Way ever experienced.}
\end{figure*}
Across all six relations shown in \autoref{fig:f6}, there is no strong dependence on the GC membership permutation. As before, Sequoia shows the most pronounced variation, but this falls within the scatter of the relations observed at $z=0$. The robustness of these results has two main implications. First, the inferred stellar masses and accretion redshifts of the satellite progenitors obtained here produce scaling relations that are in satisfactory agreement with observations at $z=0$, lending some further credence to the analysis presented in this work. Secondly, we find some evidence that the specific frequency $\tn$ changes with redshift, consistently with models suggesting that GC formation is fundamentally a baryonic process that results in relations with host galaxy properties that evolve in the context of hierarchical galaxy formation and evolution \citep[e.g.][Bastian et al.\ in prep.]{kruijssen15b,choksi18,pfeffer18,elbadry19,kruijssen19d}.

\section{Merger history of the Milky Way} \label{sec:hist}

\subsection{Merger mass ratios}
Having inferred the accretion redshifts and the stellar masses at the time of accretion of Kraken, \ges, the progenitor of the Helmi streams, Sequoia, and Sagittarius, we place these results in the context of the formation and assembly history of the Milky Way. Because these accretion events took place at different points in the Milky Way's history, it is somewhat non-trivial to assess how major (or minor) these mergers were. To calculate the merger mass ratios, we combine the stellar masses and accretion redshifts of the five satellite progenitors with the stellar mass growth history of the Milky Way that we inferred in \citet[fig.~4]{kruijssen19e}. We use a Monte Carlo approach to account for the uncertainties on this mass growth history. The mass growth history results from comparing the age-metallicity distribution of in-situ GCs, formed in the Main progenitor of the Milky Way, to the age-metallicity-mass distribution of central galaxies in the \eagle simulations. Both the mass growth history and its corresponding chemical enrichment history are consistent with independent constraints inferred from the chemical abundances of thick disc field stars and the resulting star formation history \citep[see \citealt{kruijssen19e} for further discussion]{snaith14,snaith15}. The merger mass ratio then follows as $\ms/M_{\rm \star,MW}(\zacc)$, where $M_{\rm \star,MW}(\zacc)$ is the stellar mass of the Milky Way at the time of accretion. By following this procedure, we obtain a predicted mass ratio for each of the 10,000 Monte Carlo realisations of the neural network and for each GC membership permutation.

\begin{figure*}
\includegraphics[width=\hsize]{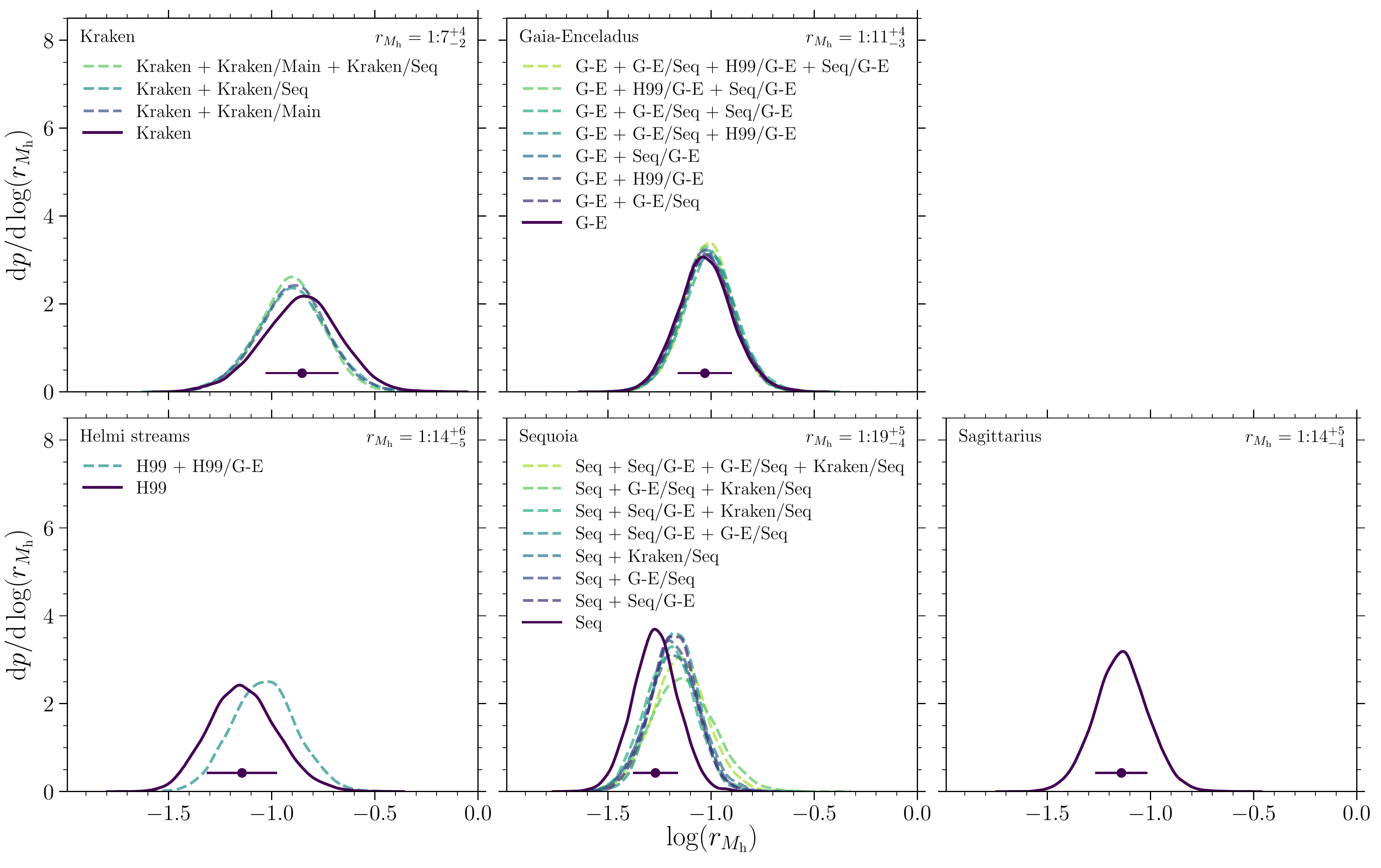}%
\caption{
\label{fig:f8}
PDFs of the predicted merger halo mass ratios of the satellite progenitor accretion events onto the Milky Way, inferred by applying a neural network trained on the \emosaics simulations to their GC populations and adopting the relation between stellar mass and halo mass at the accretion redshift from \citet{moster13} to obtain the halo masses. Each PDF shows the distribution across all 10,000 Monte Carlo realisations of the neural network, applied to different GC membership permutations as indicated by the legend (also see Table~\ref{tab:member}). In each panel, the data point with error bar and the annotation in the top right indicate the median and 16th-to-84th percentiles for the experiment using the unambiguous member GCs (corresponding to the solid line in that panel). This figure shows that the five accretion events are all minor mergers, spanning a wide range of mass ratios $\rhmass=1$:$(5{-}24)$. The accretion of Kraken likely represents the most major merger that the Milky Way ever experienced.}
\end{figure*}
\autoref{fig:f7} shows the PDFs of the merger stellar mass ratio, i.e.\ the ratio between the stellar mass of the satellite progenitors at the time of accretion (see Section~\ref{sec:sims} for its definition) and the stellar mass of the Milky Way at that time. Because the stellar mass ratio is a derived quantity that combines several variables with their own uncertainties (satellite stellar mass, accretion redshift, Milky Way stellar mass), the uncertainties on the mass ratios are considerable. None the less, \autoref{fig:f7} reveals a wide range of mass ratios, even if all accretion events represent minor mergers, i.e.\ with mass ratios $\rmass<1$:$4$. Kraken is very likely to have been the most major merger that the Milky Way ever experienced,\footnote{Of course, the Milky Way may have experienced very early major mergers with galaxies hosting few or no GCs. In Section~\ref{sec:growth}, we suggest that any such mergers must have had masses considerably smaller than $\ms\sim10^8~\msun$, suggesting that the Milky Way would have had a mass of $\ms\la10^9~\msun$. For the stellar mass growth history of the Milky Way that we inferred in \citet{kruijssen19e}, this would imply merger redshifts of $z\ga4$. These redshifts were characterised by such high merger rates \citep[with major merger rates of $\sim1~\gyr^{-1}$, see e.g.][]{fakhouri10} that it would be challenging to identify discrete events. We therefore maintain the statement made here and consider the accretion history of the Milky Way prior to $z\sim4$ to have been `a mess'.} with a mass ratio of $\rmass=1$:$31^{+34}_{-16}$. This mass ratio is well in excess of that of \ges ($\rmass=1$:$67^{+41}_{-27}$), because Kraken accreted much earlier, when the stellar mass of the Milky Way was a factor of $\sim3$ lower than when \ges accreted. Inspection of the PDF across all 10,000 Monte Carlo realisations shows that Kraken has a probability of 0.3~per~cent to have been a major merger, with a mass ratio of $\rmass>1$:$4$. For all other satellite progenitors, this probability is $<0.01$~per~cent -- we can only provide an upper limit, because among the 10,000 Monte Carlo realisations there are no cases of a mass ratio $\rmass>1$:$4$. The other accretion events, i.e.\ of the progenitor of the Helmi streams, Sequoia, and Sagittarius, are truly minor, with mass ratios of $\rmass=1$:$110^{+99}_{-54}$, $\rmass=1$:$191^{+100}_{-67}$, and $\rmass=1$:$104^{+70}_{-43}$, respectively. This implies that the Milky Way must have grown in mass mostly by gas accretion and in-situ star formation \citep[also see][]{trujillogomez20}.

For Kraken and \ges, the choice of GC membership permutation affects the merger mass ratio by 0.1~dex or less,\footnote{This spread is smaller than the corresponding spread of satellite stellar mass in \autoref{fig:f4}, because the stellar mass and accretion redshift are covariant. For lower accretion redshifts (and a more massive Milky Way at the time of accretion), the model also predicts a higher satellite mass, which suppresses the variation of the merger mass ratio.} highlighting that these mass ratios are robust. However, the progenitor of the Helmi streams and (especially) Sequoia have merger mass ratios that are quite sensitive to membership selection. If NGC5634 and NGC5904 (representing the `H99/G-E' group in Table~\ref{tab:member}) are included, the merger mass ratio of the progenitor of the Helmi streams becomes $\rmass=1$:$70^{+59}_{-33}$, due to the combination of a higher accretion redshift and a higher stellar mass. Likewise, adding any of the ambiguous GCs to Sequoia increases its merger mass ratio, which for most membership permutations also results from a combination of a higher accretion redshift and a higher stellar mass. In all cases including at least one group of ambiguous GCs, Sequoia's mass ratio increases to around $\rmass\approx1$:$130$. While these represent genuine systematic uncertainties, the qualitative conclusion remains unchanged -- both the progenitor of the Helmi streams and Sequoia represent truly minor mergers.

\begin{table*}
  \caption{Summary of the inferred properties of the satellite progenitors, for all GC membership permutations (see Table~\ref{tab:member} for the GC memberships). For each progenitor the first row lists the results obtained using the GCs that are unambiguous members. From left to right, the columns list the progenitor (and GC membership permutation), the number of associated GCs, the accretion redshift, log stellar mass, log halo mass, specific frequency, log number of GCs per unit halo mass, log total GC system mass per unit halo mass, the stellar mass ratio of the merger with the Milky Way, and the halo mass ratio of the merger with the Milky Way (determined indirectly using the relation between stellar mass and halo mass from \citealt{moster13}, see the text).}
\label{tab:summary}
  \begin{tabular*}{\textwidth}{l @{\extracolsep{\fill}} c c c c c c c c c}
   \hline
   Progenitor & $\ngc$ & $\zacc$ & $\log \ms$ & $\log \mh$ & $\tn$ & $\log \etan$ & $\log \etam$ & $r_{M_\star}$ & $r_{M_{\rm h}}$ \\ 
    &  &  & $[\msun]$ & $[\msun]$ & $[10^{-9}~{\rm M}_\odot^{-1}]$ & $[{\rm M}_\odot^{-1}]$ & & & \\ 
   \hline
   Kraken & $13$ & $2.26^{+0.39}_{-0.45}$ & $8.28^{+0.18}_{-0.17}$ & $10.92^{+0.10}_{-0.10}$ & $69^{+33}_{-23}$ & $-9.81^{+0.10}_{-0.10}$ & $-4.50^{+0.10}_{-0.10}$ & $1$:$31^{+34}_{-16}$ & $1$:$7^{+4}_{-2}$ \\ 
   Kraken + Kraken/Main & $14$ & $1.88^{+0.37}_{-0.37}$ & $8.35^{+0.18}_{-0.19}$ & $10.95^{+0.10}_{-0.10}$ & $63^{+34}_{-21}$ & $-9.81^{+0.10}_{-0.10}$ & $-4.36^{+0.10}_{-0.10}$ & $1$:$39^{+36}_{-18}$ & $1$:$8^{+4}_{-2}$ \\ 
   Kraken + Kraken/Seq & $14$ & $2.17^{+0.34}_{-0.40}$ & $8.22^{+0.16}_{-0.16}$ & $10.89^{+0.09}_{-0.09}$ & $85^{+37}_{-26}$ & $-9.74^{+0.09}_{-0.09}$ & $-4.46^{+0.09}_{-0.09}$ & $1$:$39^{+40}_{-19}$ & $1$:$8^{+4}_{-3}$ \\ 
   Kraken + Kraken/Main + Kraken/Seq & $15$ & $1.82^{+0.32}_{-0.33}$ & $8.36^{+0.16}_{-0.17}$ & $10.96^{+0.09}_{-0.09}$ & $66^{+31}_{-21}$ & $-9.78^{+0.09}_{-0.09}$ & $-4.36^{+0.09}_{-0.09}$ & $1$:$40^{+34}_{-18}$ & $1$:$8^{+4}_{-2}$ \\ 
   G-E & $20$ & $1.35^{+0.26}_{-0.23}$ & $8.43^{+0.15}_{-0.16}$ & $10.98^{+0.07}_{-0.08}$ & $75^{+32}_{-21}$ & $-9.68^{+0.08}_{-0.07}$ & $-4.27^{+0.08}_{-0.07}$ & $1$:$67^{+41}_{-27}$ & $1$:$11^{+4}_{-3}$ \\ 
   G-E + G-E/Seq & $21$ & $1.34^{+0.25}_{-0.22}$ & $8.45^{+0.14}_{-0.15}$ & $10.99^{+0.07}_{-0.08}$ & $74^{+31}_{-21}$ & $-9.67^{+0.08}_{-0.07}$ & $-4.06^{+0.08}_{-0.07}$ & $1$:$63^{+39}_{-25}$ & $1$:$11^{+4}_{-3}$ \\ 
   G-E + H99/G-E & $22$ & $1.35^{+0.24}_{-0.21}$ & $8.44^{+0.13}_{-0.14}$ & $10.98^{+0.07}_{-0.07}$ & $81^{+32}_{-21}$ & $-9.64^{+0.07}_{-0.07}$ & $-4.23^{+0.07}_{-0.07}$ & $1$:$66^{+39}_{-26}$ & $1$:$11^{+3}_{-3}$ \\ 
   G-E + Seq/G-E & $22$ & $1.36^{+0.25}_{-0.22}$ & $8.44^{+0.14}_{-0.15}$ & $10.99^{+0.07}_{-0.08}$ & $79^{+32}_{-22}$ & $-9.64^{+0.08}_{-0.07}$ & $-4.26^{+0.08}_{-0.07}$ & $1$:$63^{+40}_{-25}$ & $1$:$10^{+4}_{-3}$ \\ 
   G-E + G-E/Seq + H99/G-E & $23$ & $1.34^{+0.23}_{-0.20}$ & $8.48^{+0.13}_{-0.14}$ & $11.00^{+0.07}_{-0.07}$ & $77^{+29}_{-19}$ & $-9.64^{+0.07}_{-0.07}$ & $-4.05^{+0.07}_{-0.07}$ & $1$:$61^{+37}_{-23}$ & $1$:$10^{+3}_{-3}$ \\ 
   G-E + G-E/Seq + Seq/G-E & $23$ & $1.34^{+0.25}_{-0.22}$ & $8.48^{+0.14}_{-0.15}$ & $11.01^{+0.07}_{-0.08}$ & $76^{+31}_{-21}$ & $-9.64^{+0.08}_{-0.07}$ & $-4.07^{+0.08}_{-0.07}$ & $1$:$59^{+37}_{-23}$ & $1$:$10^{+3}_{-3}$ \\ 
   G-E + H99/G-E + Seq/G-E & $24$ & $1.31^{+0.23}_{-0.20}$ & $8.48^{+0.13}_{-0.14}$ & $11.00^{+0.07}_{-0.07}$ & $80^{+31}_{-21}$ & $-9.62^{+0.07}_{-0.07}$ & $-4.23^{+0.07}_{-0.07}$ & $1$:$63^{+37}_{-24}$ & $1$:$10^{+3}_{-3}$ \\ 
   G-E + G-E/Seq + H99/G-E + Seq/G-E & $25$ & $1.26^{+0.21}_{-0.19}$ & $8.53^{+0.12}_{-0.13}$ & $11.03^{+0.06}_{-0.07}$ & $74^{+27}_{-18}$ & $-9.63^{+0.07}_{-0.06}$ & $-4.06^{+0.07}_{-0.06}$ & $1$:$60^{+33}_{-22}$ & $1$:$10^{+3}_{-2}$ \\ 
   H99 & $5$ & $1.75^{+0.42}_{-0.37}$ & $7.96^{+0.19}_{-0.18}$ & $10.74^{+0.10}_{-0.10}$ & $54^{+29}_{-19}$ & $-10.04^{+0.10}_{-0.10}$ & $-4.71^{+0.10}_{-0.10}$ & $1$:$110^{+99}_{-54}$ & $1$:$14^{+6}_{-5}$ \\ 
   H99 + H99/G-E & $7$ & $1.93^{+0.35}_{-0.35}$ & $8.06^{+0.16}_{-0.16}$ & $10.80^{+0.09}_{-0.09}$ & $61^{+27}_{-19}$ & $-9.95^{+0.09}_{-0.09}$ & $-4.58^{+0.09}_{-0.09}$ & $1$:$70^{+59}_{-33}$ & $1$:$11^{+5}_{-3}$ \\ 
   Seq & $3$ & $1.46^{+0.17}_{-0.17}$ & $7.90^{+0.11}_{-0.11}$ & $10.70^{+0.06}_{-0.06}$ & $38^{+11}_{-8}$ & $-10.22^{+0.06}_{-0.06}$ & $-5.18^{+0.06}_{-0.06}$ & $1$:$191^{+100}_{-67}$ & $1$:$19^{+5}_{-4}$ \\ 
   Seq + Seq/G-E & $5$ & $1.52^{+0.17}_{-0.17}$ & $8.02^{+0.10}_{-0.10}$ & $10.77^{+0.05}_{-0.05}$ & $48^{+12}_{-10}$ & $-10.07^{+0.05}_{-0.05}$ & $-4.99^{+0.05}_{-0.05}$ & $1$:$132^{+69}_{-45}$ & $1$:$15^{+4}_{-3}$ \\ 
   Seq + G-E/Seq & $4$ & $1.46^{+0.20}_{-0.19}$ & $8.06^{+0.11}_{-0.11}$ & $10.79^{+0.06}_{-0.06}$ & $35^{+10}_{-8}$ & $-10.18^{+0.06}_{-0.06}$ & $-4.22^{+0.06}_{-0.06}$ & $1$:$134^{+76}_{-48}$ & $1$:$15^{+5}_{-4}$ \\ 
   Seq + Kraken/Seq & $4$ & $1.69^{+0.23}_{-0.24}$ & $7.92^{+0.11}_{-0.11}$ & $10.72^{+0.06}_{-0.06}$ & $48^{+14}_{-11}$ & $-10.12^{+0.06}_{-0.06}$ & $-5.18^{+0.06}_{-0.06}$ & $1$:$131^{+87}_{-52}$ & $1$:$15^{+5}_{-4}$ \\ 
   Seq + Seq/G-E + G-E/Seq & $6$ & $1.37^{+0.17}_{-0.15}$ & $8.14^{+0.11}_{-0.11}$ & $10.83^{+0.06}_{-0.06}$ & $43^{+12}_{-10}$ & $-10.05^{+0.06}_{-0.06}$ & $-4.23^{+0.06}_{-0.06}$ & $1$:$126^{+61}_{-41}$ & $1$:$15^{+4}_{-3}$ \\ 
   Seq + Seq/G-E + Kraken/Seq & $6$ & $1.71^{+0.20}_{-0.21}$ & $7.87^{+0.10}_{-0.10}$ & $10.69^{+0.05}_{-0.05}$ & $81^{+20}_{-17}$ & $-9.91^{+0.05}_{-0.05}$ & $-4.90^{+0.05}_{-0.05}$ & $1$:$143^{+88}_{-56}$ & $1$:$16^{+5}_{-4}$ \\ 
   Seq + G-E/Seq + Kraken/Seq & $5$ & $1.69^{+0.33}_{-0.32}$ & $8.00^{+0.15}_{-0.15}$ & $10.76^{+0.08}_{-0.08}$ & $50^{+20}_{-14}$ & $-10.06^{+0.08}_{-0.08}$ & $-4.19^{+0.08}_{-0.08}$ & $1$:$109^{+90}_{-51}$ & $1$:$14^{+6}_{-4}$ \\ 
   Seq + Seq/G-E + G-E/Seq + Kraken/Seq & $7$ & $1.56^{+0.25}_{-0.23}$ & $8.05^{+0.13}_{-0.13}$ & $10.79^{+0.07}_{-0.07}$ & $62^{+22}_{-16}$ & $-9.94^{+0.07}_{-0.07}$ & $-4.19^{+0.07}_{-0.07}$ & $1$:$115^{+77}_{-48}$ & $1$:$14^{+5}_{-4}$ \\ 
   Sagittarius & $7$ & $0.76^{+0.22}_{-0.19}$ & $8.44^{+0.22}_{-0.21}$ & $10.94^{+0.11}_{-0.10}$ & $25^{+16}_{-10}$ & $-10.10^{+0.10}_{-0.11}$ & $-4.34^{+0.10}_{-0.11}$ & $1$:$104^{+70}_{-43}$ & $1$:$14^{+5}_{-4}$ \\ 
   \hline
  \end{tabular*} 
\end{table*}
There are no estimates of the merger mass ratio available in the literature for any of the satellite progenitors considered in this work, except for \ges. \citet{helmi18} estimate that the mass ratio of the \ges accretion event was $\rmass\approx1$:$17$, which is about a factor of 4 higher than predicted by our analysis. Two ingredients contribute roughly equally to this difference. First, we predict a mass of \ges itself that is about a factor of 2.2 lower than quoted by \citet{helmi18}. We discuss in Section~\ref{sec:ms} why we expect our estimate to be more accurate. Secondly, we predict that the mass of the Milky Way at the time of accretion was about a factor of 1.8 higher than the mass adopted by \citet{helmi18}. Given the approximate nature of the numbers quoted by \citet{helmi18}, we consider our merger mass ratio to be in rough agreement, but also expect it to be more accurate, because it is based on the self-consistent orbital and chemical evolution of satellite galaxies in hydrodynamical cosmological simulations.

To complement \autoref{fig:f7}, \autoref{fig:f8} shows the PDFs of the merger halo mass ratio, i.e.\ the ratio between the halo mass of the satellite progenitors at the time of accretion (see Section~\ref{sec:sims} for its definition) and the halo mass of the Milky Way at that time (again obtained using the relation between stellar mass, halo mass, and redshift from \citealt{moster13}). Despite the fact that the halo mass ratio is a derived quantity that combines several variables with their own uncertainties, the relative uncertainties on the halo mass ratio are smaller than those on the stellar mass ratio, because the halo mass is a sub-linear function of the stellar mass. As a result, \autoref{fig:f8} reveals a relatively narrow range of halo mass ratios, with $\rhmass=1$:$(5{-}24)$ and most halo mass ratios being around $\rhmass\approx1$:$10$. The halo mass ratio obtained for \ges ($\rhmass=1$:$11^{+4}_{-3}$) is considerably smaller than that estimated by \citet{helmi18}, who find $\rhmass\approx1$:$4$. The reason is the same as for the stellar mass ratios above -- we obtain masses of both \ges and the Milky Way at the time of accretion that are both somewhat lower. In part, the discrepancy arises because \citet{helmi18} assume that \ges was similar in stellar mass to the Large Magellanic Cloud, which we find is an overestimation.

\subsection{Satellite accretion history and in-situ growth} \label{sec:growth}
Table~\ref{tab:summary} summarises the properties of the five satellite progenitors that we have quantified in this work, including each possible GC membership permutation. We also include the numbers of GCs, but point out that these are necessarily lower limits, because there is no guarantee that any of the groups are complete. All quantities listed in the table are evaluated at the time of accretion, corresponding to the accretion redshift listed in each row.

In \citet{kruijssen19e}, we estimated a total of $15\pm3$ accretion events with stellar masses $\ms>4.5\times10^6~\msun$. Even though we only discuss five such events in this paper, it is likely that these represent the most massive galaxies ever accreted by the Milky Way. Accordingly, the Milky Way never experienced a major merger since reaching a mass of $\ms\sim10^9~\msun$, which it had already attained at $z=4$ \citep{snaith14,kruijssen19e}. The Milky Way may have experienced a major merger before that time, but it is questionable how useful the concept of a major merger is when only up to a few per~cent of the current Milky Way's stellar mass could have participated.

Summing the stellar masses of the satellite progenitors at the time of their accretion yields a total accreted stellar mass of $\log{(M_{\rm \star,tot}/\msun)}=9.0\pm0.1$, which is similar in mass to the stellar halo ($1.4\pm0.4\times10^9~\msun$, see \citealt{deason19}). The remaining 10 satellites are expected to have masses below $\ms\sim10^8~\msun$ (or otherwise they would have contributed $\ga5$ GCs, which would plausibly have been identified), implying that the total accreted stellar mass is likely to be close to the sum of the five satellites discussed here (which is on the low side of, but consistent with the predictions for Milky Way-mass galaxies by \citealt{behroozi19} and \citealt{moster19}, who find $1{-}5\times10^9~\msun$). Therefore, the Galactic stellar halo can comfortably accommodate the total accreted satellite population, especially because some of the accreted mass is expected to contribute to the bulge and thick disc. In a broader context, this result means that the vast majority of the Milky Way's stellar mass formed in-situ, with only a few per~cent having been accreted in the form of stars from satellite galaxies. By contrast, the five satellite progenitors characterised here alone already contributed a total halo mass of $\log{(M_{\rm h,tot}/\msun)}=11.57\pm0.04$. This represents 25--45~per~cent of the Milky Way's halo virial mass \citep[$1.1\pm0.2\times10^{12}~\msun$, see][]{cautun20} and shows that the satellite progenitors are important contributors to the Milky Way's virial mass.

\begin{figure*}
\includegraphics[width=\hsize]{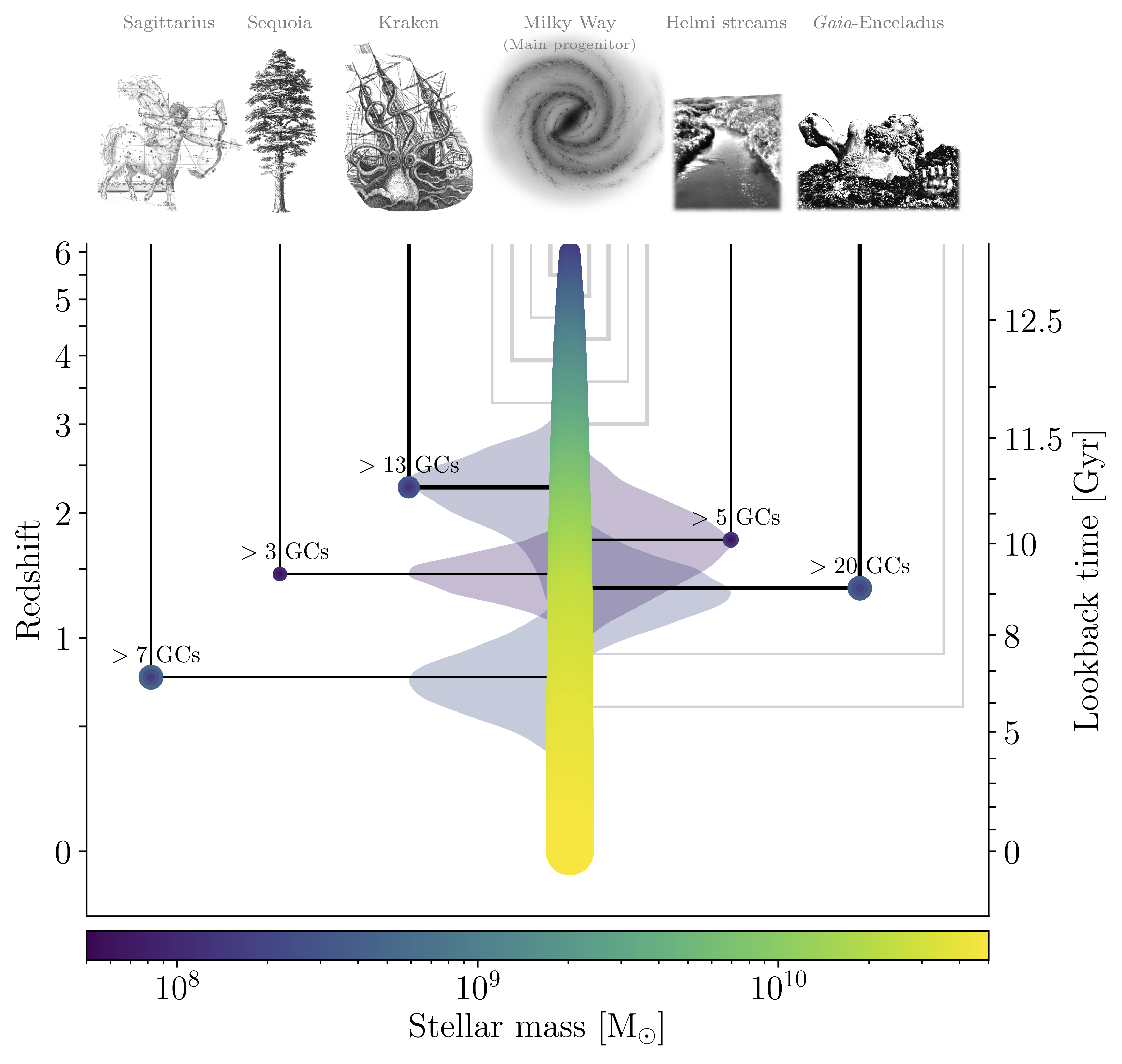}%
\caption{
\label{fig:f9}
Galaxy merger tree of the Milky Way, inferred by applying the insights gained from the \emosaics simulations to the Galactic GC population. This figure summarises many of the results presented in this paper. The main progenitor is denoted by the trunk of the tree, coloured by stellar mass (based on the reconstruction from \citealt{kruijssen19e}; see the colour bar). Black lines indicate the five identified (and likely most massive) satellites, with the shaded areas visualising the PDFs of the accretion redshifts from \autoref{fig:f5}. The coloured circles indicate the stellar masses at the time of accretion (both with their colours and sizes) and the subtly different colours of the outer rings and central dots visualise the uncertainties on the stellar masses. The annotations list the minimum number of GCs brought in by each satellite. Light grey lines illustrate the global statistics of the Milky Way's merger tree inferred by \citet{kruijssen19e}, but have no absolute meaning. Thin lines mark tiny mergers (with mass ratio $\rmass<1$:$100$) and thick lines denote minor (or possibly major) mergers (with mass ratio $\rmass>1$:$100$). This merger tree is consistent with the stellar mass growth history of the Milky Way, the total number of mergers ($\nbr$), the number of high-redshift mergers ($\nbrz$), and the numbers of tiny and minor mergers ($N_{<1:100}$ and $N_{1:100-1:4}$) from \citet{kruijssen19e}, as well as with the five identified satellite progenitors discussed in this work, including their stellar masses, accretion redshifts, and GC populations. Note that only progenitors with masses $\ms>4.5\times10^6~\msun$ are included. From left to right, the six images along the top of the figure indicate the identified progenitors, i.e.\ Sagittarius, Sequoia, Kraken, the Milky Way's Main progenitor, the progenitor of the Helmi streams, and \ges.}
\end{figure*}
A similar balance can be made for the Galactic GC population. We find 55 GCs that were likely accreted as part of the five satellite progenitors. While this number is inevitably a lower limit on the true number of ex-situ GCs, we do not expect the sample to be hugely incomplete. If the scaling relations between the number of GCs and the host galaxy mass discussed in Section~\ref{sec:scaling} apply also at the accretion redshifts of the satellite progenitors (which can be disputed), we do not expect the current sample to be incomplete by more than 30~per~cent, such that the total might be as high as 80 accreted GCs. Given a total number of 157 GCs in the catalogue of \citet[2010 edition]{harris96}, this implies that 35--50~per~cent of the Galactic GCs formed ex-situ, with the complementing 50--65~per~cent having formed in-situ. This is consistent with the estimate that 40~per~cent of GCs formed ex-situ, which we arrived at using the GC age-metallicity distribution only in \citet{kruijssen19e} and also using the kinematics of the GC population in \citet{trujillogomez20}.

Many of the results of this paper are summarised visually in \autoref{fig:f9}, which provides an extensive update to the merger tree presented in \citet[fig.~6]{kruijssen19e}. Thanks to the discovery and characterisation of several new satellites with {\it Gaia} DR2 \citep[e.g.][]{koppelman19,massari19,myeong19} and the interpretative framework provided by the \emosaics simulations \citep{pfeffer18,kruijssen19d}, we now have a reconstruction of the Milky Way's merger tree that specifically includes five out of the $15\pm3$ expected satellites with stellar masses $\ms>4.5\times10^6~\msun$. This merger tree includes the stellar mass growth history of the Milky Way's Main progenitor, the stellar masses and accretion redshifts of five satellite galaxies with stellar masses $\ms\ga5\times10^7~\msun$, and the lower limits on the numbers of GCs contributed by each of these accretion events.

\section{Conclusions} \label{sec:concl}
In this paper, we have used the \emosaics simulations to quantify the stellar masses and accretion redshifts of five satellite galaxies that have been accreted by the Milky Way, as well as several other of their properties. These satellites are Kraken (previously referred to as the progenitor of the `low-energy GCs' by \citealt{massari19}), \ges, the progenitor of the Helmi streams, Sequoia, and Sagittarius. They likely represent the most massive objects that the Milky Way accreted since $z=4$. We predict their properties by training an artificial neural network on the \emosaics simulations, which follow the co-formation and co-evolution of galaxies and their GC populations. This network relates the ages, metallicities, and orbital properties of a group of GCs that formed in a common satellite progenitor to the properties of that progenitor. The conclusions of this work are as follows.
\begin{enumerate}
\item
The neural network is capable of predicting the stellar masses and accretion redshifts of the satellite progenitors to high precision, with a validation score of $0.89^{+0.06}_{-0.06}$ and a scatter of $\sigma[\log_{10}(M_{\rm \star,pred}/M_{\rm \star,true})]=0.41^{+0.05}_{-0.04}$ and $\sigma[\log_{10}(1+z_{\rm acc,pred}/1+z_{\rm acc,true})]=0.13^{+0.01}_{-0.01}$ around the true values, respectively. Omitting the GC age-metallicity information leads to considerably worse constraints on both quantities. Additionally, the accretion redshifts strongly rely on the GC orbital information. (\autoref{fig:f1})
\item
Modulo a small number of changes, we adopt the proposed GC membership of the satellite progenitors from \citet[see our Table~\ref{tab:member}]{massari19}. The five groups of GCs associated with the different satellite progenitors show clear differences in apocentre-eccentricity space and age-metallicity space. By comparing the orbital properties of GCs to those of fossil streams from \citet{bonaca20}, we find that the streams have similar orbital properties as the GCs associated with Kraken, \ges, the progenitor of the Helmi streams, and Sagittarius. The Fimbulthul stream falls in between the orbital properties spanned by the GCs associated with Kraken and \ges, suggesting that it may be the relic of a GC that formed in either of these two satellites. In age-metallicity space, the Kraken GCs have the highest metallicities at a given age, closely followed by the \ges GCs, suggesting that these two galaxies were the most massive at any moment in time, with Kraken having a slightly higher mass. (\autoref{fig:f2} and~\ref{fig:f3})
\item
By applying the neural network to the Galactic GCs associated with the five satellite progenitors, we find that the satellites span a relatively narrow stellar mass range at the time of accretion, of $\ms=(0.6{-}4.6)\times10^8~\msun$. The top end of this range is occupied by Kraken, \ges, and Sagittarius, with the latter two being the most massive. This differs from the stellar masses at any given moment in time, because the accretion redshifts of these satellites differ. By accreting later, Sagittarius was able to attain a higher mass, despite initially being considerably less massive than Kraken and \ges. The predicted stellar masses are consistent with previous literature estimates given their uncertainties,\footnote{Differences do exist relative to previous studies that estimate projected stellar masses at $z=0$ rather than considering these at the time of accretion. See Section~\ref{sec:ms} for details.} but considerably more precise. (\autoref{fig:f4})
\item
The accretion redshifts of the five satellites span a wide range, of $\zacc=0.57{-}2.65$ (corresponding to lookback times of $\tacc=5.7{-}11.3~\gyr$). Kraken was the first galaxy to be accreted, followed by the progenitor of the Helmi streams, Sequoia and \ges, and finally Sagittarius. This order explains why the stellar mass of Sagittarius at the time of accretion exceded that of Kraken, despite having a lower mass at any given age prior to the accretion of both satellites. The predicted accretion redshifts are consistent with the rough ranges available from other works in the literature. The only possible tension with previous results is the progenitor of the Helmi streams, for which dynamical models predict a much more recent accretion ($\tacc=5{-}8~\gyr$) than obtained in this work ($\tacc=10.1^{+0.7}_{-0.9}~\gyr$), whereas the stellar age range of $\tau=11{-}13~\gyr$ is consistent with our prediction. Reconciling these predictions is an important point of attention for future work. (\autoref{fig:f5})
\item
The presented results are generally not sensitive to the details of the GC membership. By including and excluding GCs with ambiguous memberships, we find that the stellar masses and accretion redshift of the satellite progenitors exhibit differences smaller than the uncertainties estimated from the neural network. Only when including GCs that are highly unlikely to have been a members of a satellite progenitor do the results change by more than the uncertainties. (\autoref{fig:f4} and~\ref{fig:f5} and Table~\ref{tab:summary})
\item
Taking together the above results, the `low-energy' group of GCs identified by \citet{massari19} has the properties predicted for Kraken by \citet{kruijssen19e}, i.e.\ having a high mass and contributing a large number of GCs, similar to that of \ges. This high mass is required by the position of the Kraken GCs in age-metallicity space, where they occupy the high-metallicity side of the `satellite branch'. In addition, the small apocentre radii of the Kraken GCs ($\ra<7~\kpc$) requires either a high progenitor mass or a very high accretion redshift \citep{pfeffer20}. The fact that the ages of the Kraken GCs reach down to $z\sim2$ rules out the latter option. The only alternative that remains is a satellite progenitor that at any given time had a mass similar to (or higher than) \ges, as originally predicted. Kraken may represent the tip of the iceberg of several further accretion events hidden towards the Galactic centre \citep{helmi20}, where the identification of substructure is complicated by high stellar densities and phase space mixing on short orbital time-scales.
\item
We calculate the number and total mass of GCs per unit halo mass in each satellite progenitor at the time of its accretion, finding that these are consistent with the typical numbers found for the GC populations of galaxies at $z=0$. However, the specific frequencies of the Kraken and \ges exceed those of galaxies of a similar mass at $z=0$, suggesting that the scaling relations between the number of GCs and the host galaxy mass evolve with redshift \citep[e.g.][Bastian et al.\ in prep.]{kruijssen15b,choksi19}. We propose that the excess can be explained by the continued stellar mass growth of galaxies after $\zacc$. As a result, we predict that future observations of GC populations at $z>1$ should find higher specific frequencies than at $z=0$. (\autoref{fig:f6})
\item
By combining the stellar masses and accretion redshifts of the satellite progenitors with the inferred stellar mass growth history of the Milky Way's Main progenitor, we infer the stellar mass ratios of the five accretion events. We find that all accretion events are minor mergers (i.e.\ with stellar mass ratios $\rmass<1$:$4$). Thanks to its high mass and accretion redshift, Kraken is the most major merger (i.e.\ the merger with the highest mass ratio) out of the five satellites considered here, with a mass ratio of $\rmass=1$:$31^{+34}_{-16}$, making it very likely to be the most significant merger ever experienced by the Milky Way. Based on the full PDF obtained from the neural network, the accretion of Kraken by the Milky Way has a chance of 0.3~per~cent to have been a major merger event (for which $\rmass>1$:$4$). For all other satellite progenitors, the merger mass ratios are smaller than that of Kraken by factors of 2--6. This implies that the Milky Way must have grown its stellar mass mostly by gas accretion and in-situ star formation. (\autoref{fig:f7})
\item
Likewise, we find low merger halo mass ratios, in the range $\rhmass=1$:$(5{-}24)$, again indicating that all accretion events were minor mergers. With a halo mass ratio of $\rhmass=1$:$7^{+4}_{-2}$, Kraken may have been the last merger to have significantly disrupted the Galactic disc (at $\zacc=2.26^{+0.39}_{-0.45}$ or $\tacc=10.9^{+0.4}_{-0.7}~\gyr$), as all subsequent mergers (including \ges) had halo mass ratios $\rhmass<1$:$10$. (\autoref{fig:f8})
\item
We tabulate all quantities inferred for the satellite progenitors in Table~\ref{tab:summary}. The sum of their stellar masses is $\log{(M_{\rm \star,tot}/\msun)}=9.0\pm0.1$, similar to the mass of the Galactic stellar halo ($1.4\pm0.4\times10^9~\msun$, see \citealt{deason19}). We find that the stellar halo can accommodate the total accreted satellite population, and that only a few per~cent of the Milky Way's stellar mass was accreted in the form of dwarf galaxies. The same applies for GCs, for which we estimate that 35--50~per~cent has an ex-situ origin. In contrast with the stellar mass balance, the five satellite progenitors did contribute a significant total halo mass of $\log{(M_{\rm h,tot}/\msun)}=11.57\pm0.04$, which is 25--45~per~cent of the Galactic virial mass ($1.1\pm0.2\times10^{12}~\msun$, see \citealt{cautun20}).
\item
We combine the results of our analysis with those from \citet{kruijssen19e} to present the most detailed reconstruction to date of the Milky Way's merger tree. This merger tree includes a total $15\pm3$ expected satellites with stellar masses $\ms>4.5\times10^6~\msun$, as well as the stellar mass growth history of the Milky Way's Main progenitor, the stellar masses and accretion redshifts of the five satellite galaxies considered here, and the lower limits on the numbers of GCs contributed by these accretion events. (\autoref{fig:f9})
\end{enumerate}
The above results represent an example of how the phase space clustering of stellar populations in the revolutionary {\it Gaia} data can unlock a wealth of information when interpreted using state-of-the-art numerical simulations of galaxy formation and evolution. The machine learning approach employed here provides a very promising way of achieving this. By applying our formalism to the complete \emosaics volume including all galaxies in a (34~comoving~Mpc)$^3$ region, we aim to further boost its precision and diagnostic power through greatly improved statistics and a correspondingly larger training data set. This will be a critical step, because increasing the size of the training set allows the use of observables that are less information-rich. Therefore, it is to be expected that the expansion of our formalism to a larger suite of simulated galaxies will enable reconstructing the assembly histories of galaxies beyond the Milky Way, for which less comprehensive diagnostics are available. While this paper provides an application to only a single galaxy, it is clear that the field has entered an era in which GCs are now firmly established, quantitative tracers of galaxy formation and assembly.

Focusing on the assembly history of the Milky Way, our results add to a growing body of evidence that the Milky Way experienced an unusual path to adolescence. Not only did it assemble unusually quickly for its mass, but it also experienced a striking paucity of major accretion events, with only a handful of minor mergers shaping the Galactic stellar halo. The fact that it grew most of its stellar mass through secular processes and in-situ star formation implies that it may not be the most representative example for understanding the evolution and assembly of the galaxy population, but is a correspondingly more pleasant environment to live in.

\section*{Acknowledgements}
We thank Bill Harris for providing the data from \citet{harris17} in electronic form. We thank Charlie Conroy, Alis Deason, Benjamin Keller, Davide Massari, and Rohan Naidu for helpful discussions. JMDK and MC gratefully acknowledge funding from the Deutsche Forschungsgemeinschaft (DFG, German Research Foundation) through an Emmy Noether Research Group (grant number KR4801/1-1) and the DFG Sachbeihilfe (grant number KR4801/2-1). JMDK, STG, and MRC gratefully acknowledge funding from the European Research Council (ERC) under the European Union's Horizon 2020 research and innovation programme via the ERC Starting Grant MUSTANG (grant agreement number 714907). JMDK gratefully acknowledges funding from the DFG via the SFB 881 ``The Milky Way System'' (subproject B2). JP and NB gratefully acknowledge funding from the ERC under the European Union's Horizon 2020 research and innovation programme via the ERC Consolidator Grant Multi-Pop (grant agreement number 646928). MRC is supported by a Fellowship from the International Max Planck Research School for Astronomy and Cosmic Physics at the University of Heidelberg (IMPRS-HD). NB and RAC are Royal Society University Research Fellows. This work used the DiRAC Data Centric system at Durham University, operated by the Institute for Computational Cosmology on behalf of the STFC DiRAC HPC Facility (\url{www.dirac.ac.uk}). This equipment was funded by BIS National E-infrastructure capital grant ST/K00042X/1, STFC capital grants ST/H008519/1 and ST/K00087X/1, STFC DiRAC Operations grant ST/K003267/1 and Durham University. DiRAC is part of the National E-Infrastructure. This study also made use of high performance computing facilities at Liverpool John Moores University, partly funded by the Royal Society and LJMU's Faculty of Engineering and Technology. This work has made use of Matplotlib \citep{hunter07}, Numpy \citep{vanderwalt11}, and Astropy \citep{astropy13}.

Credits of the images in \autoref{fig:f9} are as follows. The image of Sagittarius is adapted from Alexander Jamieson's {\it A Celestial Atlas}, 1822, plate 25. The image of Sequoia is adapted from a line engraving of a Sequoiadendron giganteum in California (1872--1873). The image of Kraken is adapted from an illustration in Pierre Denys de Montfort's {\it Histoire Naturelle, G\'en\'erale et Particuliere des Mollusques}, Vol.\ 2, 1801, pp.\ 256--257. The image of the Milky Way is adapted from the illustration by R.~Hurt, with credit NASA/JPL-Caltech/R.~Hurt (SSC/Caltech). The image of the progenitor of the Helmi Streams is an adapted picture of the Arroyo Napost\'{a} river near Bah\'{i}a Blanca, Argentina. The image of \ges is an adapted picture of Gaspard Marsy's {\it Fontaine de l'Encelade} (1675--1676) at the Mus\'{e}e national des ch\^{a}teaux de Versailles et de Trianon in Versailles, France. 

\vspace{-4mm}

\bibliographystyle{mnras}
\bibliography{mybib}

\vspace{-3mm}

\bsp

\label{lastpage}

\end{document}